\documentclass{aa}

\usepackage[varg]{txfonts}
\usepackage{natbib}
\bibpunct{(}{)}{;}{a}{}{,}
	
\usepackage{geometry}                		
\usepackage[english]{babel}
\usepackage{graphicx}	
\usepackage{refstyle}
\usepackage{amsmath}
\usepackage{pifont}
\usepackage{amssymb}
\usepackage{float}
\usepackage{mathabx}
\newcommand{\infapp}{\raisebox{-.7ex}{$\stackrel{<}{\,\sim\,}$}}
\newcommand{\vv}{\vec{v}}
\newcommand{\vu}{\vec{u}}
\newcommand{\vg}{\vec{g}}
\newcommand{\vV}{\vec{V}}
\newcommand{\ephi}{\vec{e}_\varphi}
\newcommand{\ez}{\vec{e}_z}
\newcommand{\es}{\vec{e}_s}
\newcommand{\vO}{\vec{\Omega}}
\newcommand{\od}[1]{\mbox{${\cal O}(#1)$}}
\newcommand{\BA}{Boussinesq approximation\ }
\newcommand{\BVF}{Brunt-V\"ais\"al\"a frequency\ }
\newcommand{\sth}{ \sin\theta }
\newcommand{\cth}{ \cos\theta }

\newcommand{\PR}{\mbox{$ {\cal P} $}}
\usepackage{natbib,twoopt}
\usepackage[breaklinks=true]{hyperref} 
\bibpunct{(}{)}{;}{a}{}{,}             
\makeatletter
  \newcommandtwoopt{\citeads}[3][][]{\href{http://adsabs.harvard.edu/abs/#3}%
    {\def\hyper@linkstart##1##2{}%
     \let\hyper@linkend\@empty\citealp[#1][#2]{#3}}}
  \newcommandtwoopt{\citepads}[3][][]{\href{http://adsabs.harvard.edu/abs/#3}%
    {\def\hyper@linkstart##1##2{}%
     \let\hyper@linkend\@empty\citep[#1][#2]{#3}}}
  \newcommandtwoopt{\citetads}[3][][]{\href{http://adsabs.harvard.edu/abs/#3}%
    {\def\hyper@linkstart##1##2{}%
     \let\hyper@linkend\@empty\citet[#1][#2]{#3}}}
  \newcommandtwoopt{\citeyearads}[3][][]%
    {\href{http://adsabs.harvard.edu/abs/#3}
    {\def\hyper@linkstart##1##2{}%
     \let\hyper@linkend\@empty\citeyear[#1][#2]{#3}}}
\makeatother

\begin{document}

\title{The 2D dynamics of radiative zones of low-mass stars}

\author{D. Hypolite\inst{1},
S. Mathis\inst{1}
\and M. Rieutord\inst{2}}

\institute{Laboratoire AIM Paris-Saclay, CEA/DRF - CNRS - Université Paris-Saclay, IRFU/DAp Centre de Saclay, F-91191 Gif-sur-Yvette Cedex, France
\and Institut de Recherche en Astrophysique et Planétologie, Observatoire Midi-Pyrénées, Université de Toulouse, 14 avenue Edouard Belin, 31400 Toulouse, France}

\date{\today}

\abstract{
Helioseismology and asteroseismology allow us to probe the differential rotation deep within low-mass stars.
In the solar convective envelope, the rotation varies with latitude with an equator rotating faster than the pole, which results in a shear applied on the radiative zone below. However, a polar acceleration of the convective envelope can be obtained through 3D numerical simulations in other low-mass stars and the dynamical interaction of the surface convective envelope with the radiative core needs to be investigated in the general case.
}{
In the context of secular evolution, we aim at describing the dynamics of the radiative core of low-mass stars to get a deeper understanding of the internal transport of angular momentum in such stars which results in a solid rotation in the Sun from $0.7R_\odot$ to $0.2R_\odot$ and a weak radial core-envelope differential rotation in solar-type stars. This study requires at least a 2D description to capture the latitudinal variations of the differential rotation.
}{
We build 2D numerical models of a radiative core on the top of which we
impose a latitudinal shear so as to reproduce a conical or cylindrical
differential rotation in a convective envelope. We perform a systematic
study over the Rossby number $\mathcal{R}o=\Delta\Omega/2\Omega_0$
measuring the latitudinal differential rotation at the
radiative-convective interface. We provide a 2D description of the
differential rotation and the associated meridional circulation in the
incompressible and stably stratified cases using the Boussinesq approximation.
 }{
The imposed shear generates a geostrophic flow implying a cylindrical differential rotation in the case of an isotropic viscosity. 
When compared to the baroclinic flow that arises from the stable
stratification, we find that the geostrophic flow is dominant when the
Rossby number is high enough ($\mathcal{R}o\geq 1$) with a cylindrical
rotation profile. For low Rossby numbers ($\mathcal{R}o < 1$), the
baroclinic solution dominates with a quasi-shellular rotation profile. Using scaling laws from 3D simulations, we show that slow rotators ($\Omega_0<30\Omega_\odot$) are expected to have a cylindrical rotation profile. Fast rotators ($\Omega_0>30\Omega_\odot$) may have a shellular profile at the beginning of the main-sequence in stellar radiative zones.
}{
This study enables us to predict different types of differential rotation and emphasizes the need of a new generation of 2D rotating stellar models developed in synergy with 3D numerical simulations. The shear induced by a surface convective zone has a strong impact on the dynamics of the underlying radiative zone in low-mass stars. But, it cannot produce a flat internal rotation profile in a solar configuration calling for additional processes for the transport of angular momentum in both radial and latitudinal directions.
}


\authorrunning{Hypolite, Mathis \& Rieutord}
\titlerunning{The 2D dynamics of the radiative zone of low-mass stars}

\maketitle

\section{Introduction}

Rotation is a key physical mechanism regarding the dynamical, chemical and magnetic
evolution of stars \citep{maeder09,charbonneau10,brun15}. 
Within the context of secular evolution, rotating radiative zones are particularly interesting since they impose the transport timescales of angular momentum and chemicals.
Indeed, differential rotation induces
meridional circulation and potential shear instabilities that transport chemical elements and angular momentum \citep{zahn92, maederzahn98, mathiszahn04, MR06} and impact the rotation profiles and chemical abundances.

In this context, helio- and asteroseismology have been revolutionary, probing for the first time the internal dynamical state of stars.
In the Sun, this reveals a uniform rotation profile in the radiative core \citep[see e.g. the review of ][]{thompson03} at least until $0.2R_\odot$ \citep{couvidat03, garcia07}; the estimate of differential rotation is more difficult for deep regions such as near the center of the Sun because of the lack of identification of individual
g-modes \citep{appourchaux10}.
In the convective envelope, the differential rotation is found conical with a
prograde equatorial acceleration and the azimuthal velocity decreasing
monotonically towards higher latitude.  A strong radial differential
rotation is observed at the bottom of the convective zone forming a thin
shear zone called the tachocline \citep{spiegelzahn92}.
This calls for strong transport processes operating at the radiative-convective interface that
must be investigated in order to understand the interactions between these two regions
\cite[e.g.][]{garaud02b, brun06, brun11, strugarekX, varela16}.  

The internal differential rotation of several other types
of stars has been revealed by asteroseismology.
Indeed, the core-to-surface rotation ratio of numerous low-mass, main-sequence,
subgiant and red giant stars \citep{beck12, deheuvels12, deheuvels14, benomar15, deheuvels16} provide new constraints for
stellar modeling .  
Moreover, a surface differential rotation is always observed in the convective envelope of low-mass stars \citep{barnes05, reinhold13} and it has been confirmed by numerical simulations \citep{brun15}.

In this context, 3D numerical simulations have been performed to better understand the magnetic activity of solar-like stars \citep{brun02, brun04, brown08, matt11, brun11, augustson12} and determine
what influences the sign of the latitudinal differential rotation in such stars
\citep{gastine14, brun17}. These studies, devoted to short timescales, show
that solar parameters can produce anti-solar rotation profiles (slow
equator and fast pole; e.g. \citealt{kapyla14}). Precisely, depending on the convective "fluid" Rossby number $\mathcal{R}o_f=v/2\Omega_0 R$, where $v$ is the velocity, $R$ the stellar radius and $\Omega_0$ is the stellar rotation rate, three rotational states have been identified in the convective
envelope of low-mass stars:  the anti-solar differential rotation for $\mathcal{R}o_f>1$, the solar-like profile when the convective Rossby number is between $0.3$ and $0.9$, and the "Jupiter-like" profile (cylindrical banded profile with alterning fast and slow jets) for $\mathcal{R}o_f<0.3$ \citep{brun17}.
In the solar case, the isorotational contours within the convective zone have also been
fitted with characteristics of the thermal wind equation showing a strong correlation between
the entropy and the angular velocity \citep{B09a, B09b, B10a, B09b, B11, brun11}. They show a
very good agreement with helioseismology data. However, these works emphasize their difficulty
to reproduce the solar tachocline, where the thermal wind balance breaks, and the underlying
flows in the radiative zone.

In this work, we aim to understand the impact of the shear of the convective envelope on the underlying radiative core on secular timescales along stellar evolution.
Unfortunately, global 3D numerical simulations focus 
on the dynamical timescales and are not able to explore intensively the entire H-R diagram for now. In this direction, considerable efforts have been made to
solve the rotational dynamics of stellar radiative zones in 1D
stellar evolution models \citep[e.g. ][]{talon97, meynetmaeder00, palacios03, TC05, eggenberger05, decressin09, marques13, mathis13}. 
When ignoring internal gravity waves and magnetic fields, and using the formalism by \cite{zahn92}, \cite{maederzahn98} and \cite{mathiszahn04} that assumes a shellular differential rotation enforced by a strong horizontal turbulence,
they fail to reproduce the rotation profile of the solar radiative core and the core-envelope rotation contrasts revealed by asteroseismology.
Moreover, rotating flows are intrinsically bidimensional. 
Then, the differential rotation can be radial and latitudinal and a more general 2D approach is needed.
In this context, the recent improved angular momentum evolution
models \citep{GB13, GB15, amard16}, which follow the
rotational evolution of low-mass stars in clusters, highlight the need of fast rotating models, and therefore 2D models, for example during the early evolution phases \citep{HR14}.

Indeed, in the case of fast rotation, \cite{MR06}, \cite{ELR13}, \cite{RB14} and
\cite{HR14} show that the baroclinic flow that pervades
rotating radiative envelopes exhibits a meridional circulation and
associated differential rotation that require a 2D description. 
The models they develop are intermediate-mass main-sequence stars models (a radiative envelope lying upon a convective core) calling now for low-mass star models with the convective envelope on top of the radiative core.
Regarding the solar case, \cite{friedlander76} studied the spin-down of a radiative zone due
to applied surface stresses using the Boussinesq approximation.  
\cite{garaud02} describes the meridional flows in a radiative zone submitted to an imposed
solar-like latitudinal shear, based on observations, to reproduce
the presence of the convective zone at its top using the anelastic
approximation. 
However, these works are limited
to the solar case and a general study of the dynamics of a radiative
zone lying below a convection zone with solar or anti-solar stress is
an interesting and necessary complement to the existing models.

For these reasons, we propose a new 2D study of the dynamics of the
radiative zone of low-mass stars.  Without any magnetic field as a first
step, neither convective motions and internal gravity waves, a 2D description is enough to capture
the essential of the physics of a rotating radiative zone submitted to
a shear at its upper boundary. We construct a latitudinal shear
boundary condition based on the results from inverted rotation profile
within the solar convective envelope and from numerical simulations of low-mass stars convective envelopes. The shear is quantified by the Rossby number $\mathcal{R}o=\Delta\Omega/2\Omega_0$, which is the normalized latitudinal differential rotation at the convective-radiative interface relative to the stellar rotation rate.
We produce solar and anti-solar configurations and solve for the flow of
an incompressible fluid and then for a stably stratified fluid using the
Boussinesq approximation. In section \ref{sect1},
we give a complete description of the incompressible model and the
hydrodynamical equations we solve when imposing a latitudinal shear.
We introduce the relevant physical parameters of the problem.
An analytical study unravels the dynamics of the bulk of
the radiative core. 
We describe the properties of the flow and compare the analytical solutions to numerical simulation solutions.  
In section \ref{sect2} and \ref{sect3}, we study the stratified
case using the Boussinesq approximation. 
Using 1D models of solar-like stars to compute the baroclinic torque amplitude (proportional to the Brunt-V\"ais\"al\"a frequency), we provide the 2D differential
rotation and meridional circulation varying systematically the Rossby number.
We also compute the
core-to-surface rotation ratio as a function of the Rossby number. 
In section \ref{sect3}, we summarize
our main results, namely we describe the Rossby parameter regime where the geostrophic solution arising from the shear dominates the dynamics ($\mathcal{R}o\geq 1$)
or the baroclinic flow dominates  the dynamics ($\mathcal{R}o< 1$).
Using scaling laws derived from 3D numerical simulations, we are able to scale the Rossby number and predict the rotational state of a low-mass stars radiative core induced by hydrodynamical processes as a function of its age and angular velocity.

\section{The flow driven by an imposed differential rotation}
\label{sect1}

\subsection{Equations of motion}

We consider an incompressible viscous fluid enclosed within
a sphere.  The system is rotating at
a constant rate $\vec{\Omega}_0$ aligned with the $z$-axis
($\vec{\Omega_0}=\Omega_0\vec{e}_z$) and the sphere is assumed, as a
first step, not to suffer any deformation due to rotation. The hydrostatic deformation of the sphere is neglected. Therefore,
we focus on the effects of the Coriolis acceleration.  Solutions are
axisymmetric and symmetric with respect to the equator as no mechanism
acting in this setting (gravity and rotation) can depart the velocity
field from these symmetries.  The dynamics of such a fluid is governed
by the momentum equation

\begin{equation}
\rho_0 [ \partial_t\vec{v} + (\vec{v}\cdot\vec{\nabla})\vec{v}
+2\vec{\Omega_0} \wedge \vec{v}  -\Omega_0^2s\vec{e}_s ]=-\vec{\nabla}P+ \rho_0 \vec{g} +\mu\Delta\vec{v}\; ,
\end{equation}
which we write in a frame rotating at angular velocity $\vec{\Omega}_0$.
In this equation we recognize the Coriolis acceleration $2\vec{\Omega_0}
\wedge \vec{v}$ and the centrifugal acceleration
$-\Omega_0^2s\vec{e}_s$, where $\vec{e}_s$ is the radial unit vector
associated with the radial cylindrical coordinate $s$. $P$ is the
pressure, $\vg$ the gravity and $\mu$ the dynamical viscosity.
Mass conservation implies

\begin{equation}
\vec{\nabla}\cdot\vec{v}=0\; ,
\end{equation}
for an incompressible fluid of constant density $\rho_0$.\\

We then gather the centrifugal acceleration, the pressure and the
gravitational potential $\Phi$ into a single scalar \mbox{$\Pi=\frac{P}{\rho_0}
+ \Phi -\frac{1}{2}s^2\Omega_0^2$} so that the momentum equation becomes

\begin{equation}
\partial_t\vv + (\vv\cdot\vec{\nabla})\vv + 2\vec{\Omega_0}\wedge\vv=
-\vec{\nabla}\Pi +\nu\Delta\vv\; ,
\end{equation}
where $\nu=\frac{\mu}{\rho_0}$ is the kinematic viscosity.

\subsection{Boundary conditions}
\label{bc1}

Since we wish to describe a
radiative core, the bounding sphere materializes the interface with an
outer convective envelope. To make this interface simple, we 
assume that the convective envelope imposes its azimuthal velocity at
the top of the radiative region. We neglect any other motion, and any
mass exchange. Hence at $r=R$ ($R$ is the radius of the radiative-convective interface) we impose

\begin{equation}
\vv= R \sin\theta \Omega_{cz}(r=R,\theta) \ephi\; ,
\end{equation}
with
\begin{equation}
\Omega_{cz}(r=R,\theta)= \Omega_0 + \Delta\Omega\sin^2\theta\; ,
\end{equation}
which is the simplest expression we can take inspired by numerical simulations
\citep[e.g. ][]{matt11, kapyla14}.  Differential rotation is then called ‘solar-like’ when
$\Delta\Omega>0$, i.e. the equatorial regions rotate faster than the pole,
and ‘anti-solar’ otherwise, i.e. $\Delta\Omega<0$.

In the frame corotating with the pole of the model, the boundary condition
reads

\begin{equation}
v_\varphi(r=R,\theta)= R \sin^3\theta\ \Delta\Omega; .
\label{scaledbc}
\end{equation}

For the sake of simplicity, we also assume the meridional components
of the velocity field at the upper boundary layer to be zero in the
rotating frame, namely

\begin{equation}
v_r(r=R,\theta)=v_\theta(r=R,\theta)=0 \; .
\label{scaledbcpsi}
\end{equation}
We note that these boundary conditions imply that viscous stresses are
applied on top of the radiative core, which is different from
the boundary conditions chosen by \cite{garaud02}, who assumed that the
continuity of the viscous stresses affecting the meridional circulation
imposes stress-free like boundary conditions at the interface. Besides,
\cite{brunzahn06} considered impenetrable walls at the bottom of
the convective envelope ($u_r=0$) and stress-free conditions on the
latitudinal and azimuthal components of the velocity field.

Here, we choose
to impose the velocity as if the convection zone would behave as a solid
that can absorb any stress. This is certainly exagerated and this excludes
any mass exchange between the layers. Nevertheless, if we consider that
the turbulent convection zone is endowed with a turbulent viscosity,
much larger than that of the radiative zone, the shear-stress of the
radiative zone is likely to be unimportant to modify the flow in
the convective zone. Thus imposing the velocity is likely more
appropriate than imposing the stress.

Our boundary conditions are simple and finally just assume that the flow
in the radiative zone does not feed back on the convection zone mean
flows.

\subsection{Scaled equations}

We adimensionalize the equations choosing 

\begin{itemize}
\item $R$ as a length scale,

\item $V= R|\Delta\Omega|$ as a velocity scale,

\item and $(2\Omega_0)^{-1}$ as the time scale.
\end{itemize}

\noindent It yields

\begin{multline}
\left \{
\begin{array}{lcl}
\partial_\tau\vec{u} + \mathcal{R}o~ (\vu\cdot\vec{\nabla})\vu + \ez\wedge\vu
 =-\vec{\nabla}p+E\Delta\vec{u}\; ,\\
\\
\vec{\nabla}\cdot\vec{u} = 0\; ,\\
\end{array}
\right.
\label{eq1}
\end{multline}
where $\vec{u}$, $\tau$ and $p$ are respectively the dimensionless
velocity, time and reduced pressure.  We introduced the two numbers:

\begin{equation}
\mathcal{R}o=\frac{\Delta\Omega}{2\Omega_0}, \qquad E=\frac{\nu}{2\Omega_0 R^2},
\end{equation}
namely the Rossby number $\mathcal{R}o$ and the Ekman number $E$.

We evaluate the order of magnitude of these numbers with the Sun.
We take $R\sim 0.7 R_\odot$ and a rotation rate at the solar tachocline
corresponding to a rotation period of $27$ days \citep{brun04}.
From \cite{ELR13}, we estimate the
kinematic viscosity either $\nu\sim10^{2}$~m$^{2}$.s$^{-1}$ 
or $\nu\sim10^{4}$~m$^{2}$.s$^{-1}$ if some turbulence occurs
\citep[e.g.][]{zahn92}. The resulting Ekman numbers are $E\sim10^{-10}$
and $E\sim10^{-8}$ respectively.

As far as the Rossby number is concerned, a typical value for the Sun is
$\sim0.1$ \citep[e.g. ][]{gastine14}. As a further simplification, we shall set this number to zero
so as to deal with linear equations focusing mainly on the case of rapid rotation.
This assumption is valid as long as the Coriolis term is dominant compared to the non-linear advection term which reads as the condition $u_\varphi\ll \mathcal{R}o^{-1}$. This condition is satisfied as shown in the next section as long as the Rossby number is less then one.

Taking the curl of the momentum equation, we get a simple vorticity
equation for the steady state

\begin{equation}
\vec{\nabla}\wedge(\vec{e}_z\wedge\vec{u}-E\Delta\vec{u})= 0\; .
\label{eq2}
\end{equation}
It is completed by the mass conservation equation

\begin{equation}
\vec{\nabla}\cdot\vu = 0 \; ,\label{eq3}
\end{equation} 
and the boundary condition

\begin{equation}
\vu = b\sin^3\theta\ephi \; , \quad {\rm at}\quad r=1 \; ,
\label{bcu}
\end{equation}
where the parameter $b=\pm1$ captures the sign of $\Delta\Omega$.

\subsection{Analytical solution}

The inviscid ($E=0$) case of this setting is the geostrophic balance

\begin{equation}
\vec{e}_z\wedge\vec{u} =-\vec{\nabla}p\; ,
\label{geoeq}
\end{equation}
which has for solution an azimuthal geostrophic flow whose amplitude
only depends on the radial cylindrical  coordinate $s$, namely

\begin{equation}
\bar{\vec{u}}=F(s)\vec{e}_\varphi\; ,
\end{equation}
as a consequence of Taylor-Proudman theorem.  The sum of the inviscid
solution and its boundary layer correction $\bar{\vec{u}}+\tilde{\vec{u}}$
has to match the boundary conditions (\ref{bcu}) and solves the equations
(\ref{eq2}) and (\ref{eq3}) when $E\neq 0$.

In this particular case, no boundary layer correction is necessary on
the azimuthal flow and we readily write

\begin{equation}
F(s) = bs^3\; ,
\end{equation} 
which satisfies the geostrophic balance (\ref{geoeq}) and boundary
condition (\ref{bcu}).  The meridional circulation arises from the
conservation of angular momentum expressed by the $\varphi$-component of
the momentum equation. In cylindrical coordinates it reads

\begin{equation}
u_s=E\left(\Delta-\frac{1}{s^2}\right)u_\varphi\; ,
\label{eqmeridional}
\end{equation}
and leads to the following expression of the meridional flow

\begin{equation}
\vu_{\rm merid} = 8bE(s\vec{e}_s -2z\vec{e}_z)\; ,
\label{mercirc}
\end{equation} 
where mass conservation has been used. 

The meridional stream function defined as
\mbox{$\vu_{\rm merid}=\vec{\nabla}\times[\psi(r,\theta)\vec{e_\varphi}]$} and leading to

\begin{equation}
u_r=\frac{1}{r\sin\theta}\partial_\theta (\sin\theta \psi) \quad
\text{and} \quad u_\theta=-\frac{1}{r}\partial_r(r \psi)\; ,
\end{equation}
reads

\begin{equation}
\psi=-8bEsz\; .
\end{equation}
We note that the foregoing meridional circulation does not meet the
imposed boundary conditions at $r=1$ for the radial and latitudinal velocities. Indeed, $u_r(1)\neq0$. This
implies the existence of a thin Ekman layer that absorbs this mass flux
and generate an \od{\sqrt{E}}-$\tilde{u}_\theta$ correction to the interior
$\bar{u}_\theta$ of (\ref{mercirc}).

Numerical solutions of the next section validate the foregoing
predictions on the flow.

\subsection{Numerical solutions}
\label{MN}

To prepare for the study of a more complex situation (the stably
stratified case), we now solve Eqs. (\ref{eq2}) and
(\ref{eq3}) with a spectral numerical method \citep{R87}. Briefly, we expand the velocity
fields on the vectorial spherical harmonics, and discretize the radial
functions on the Gauss-Lobatto grid associated with
Chebyshev polynomials. A more detailed description of the method is
given in the appendix \ref{MN2} of the paper.

\begin{figure}
 	\begin{center}
 \includegraphics[scale=0.17]{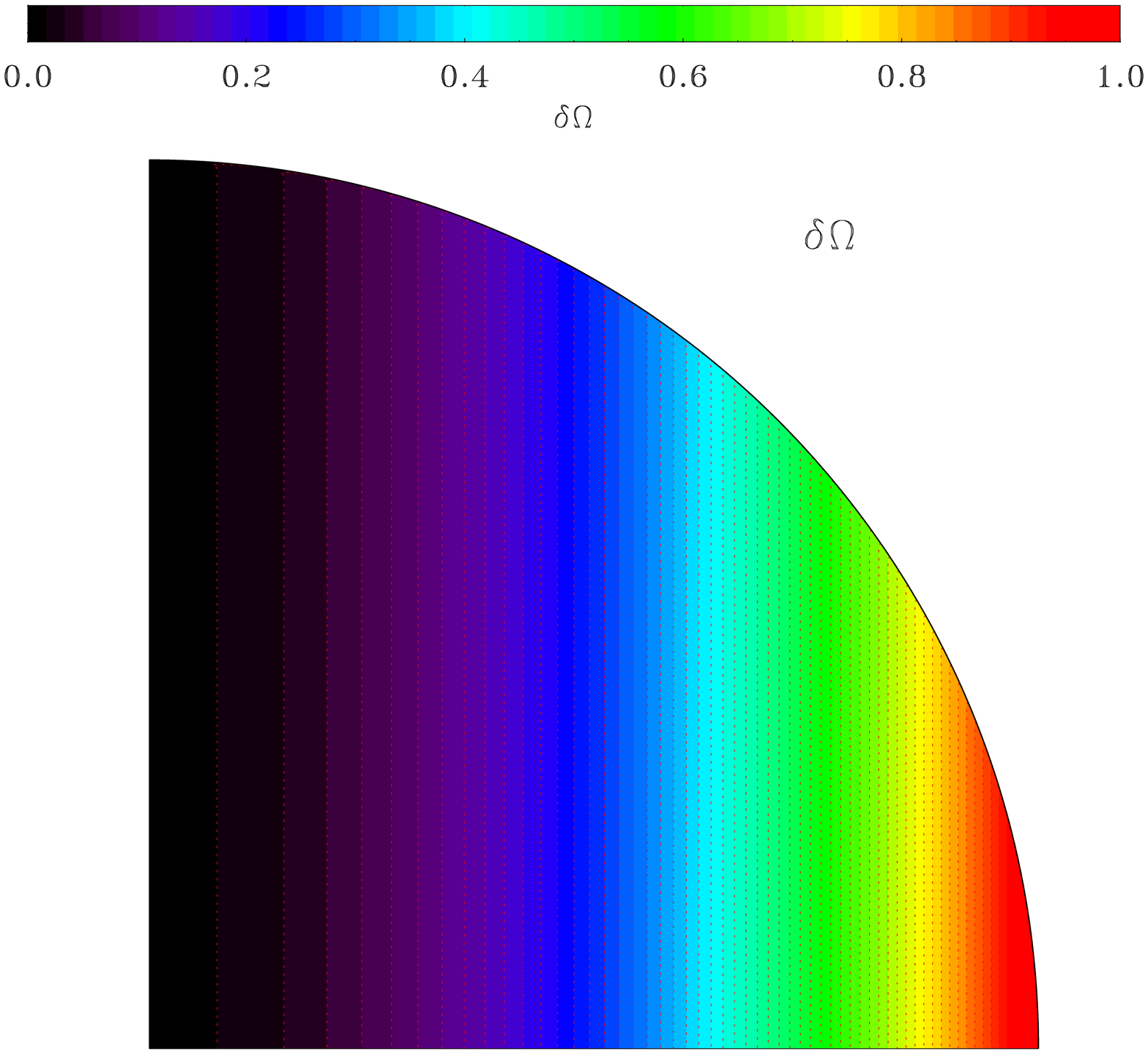}
 \includegraphics[scale=0.17]{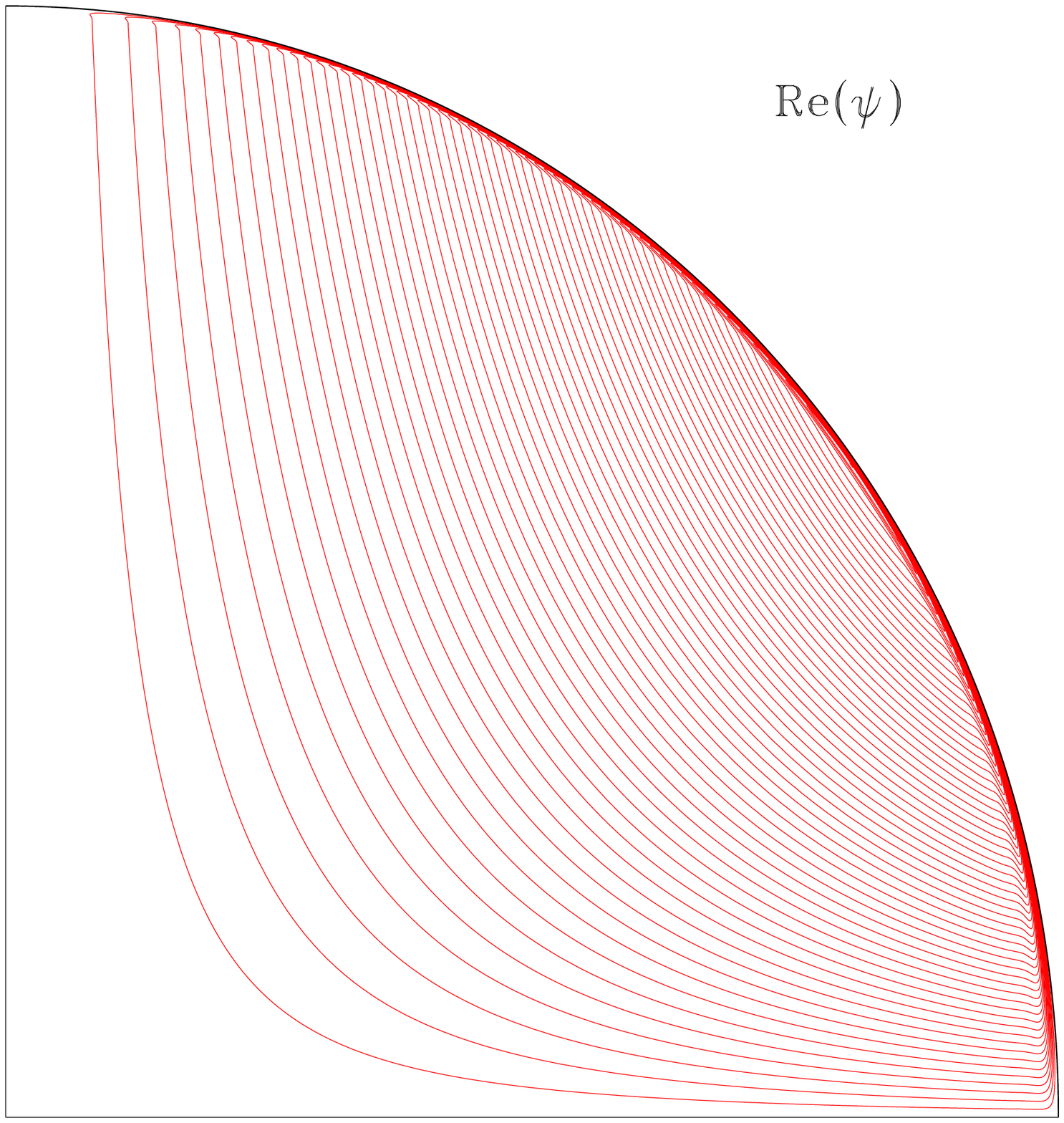}\\
 \includegraphics[scale=0.17]{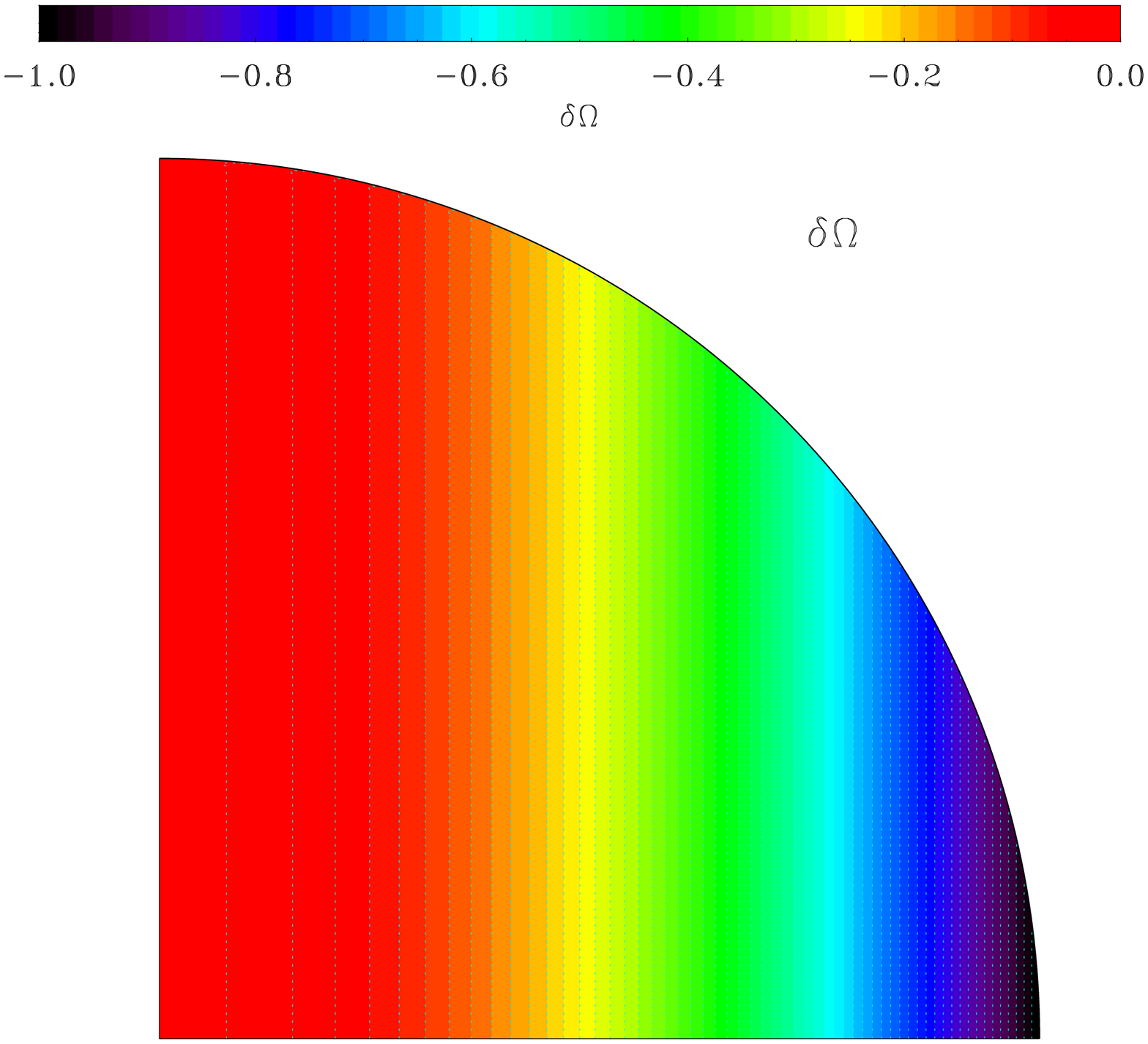}
 \includegraphics[scale=0.17]{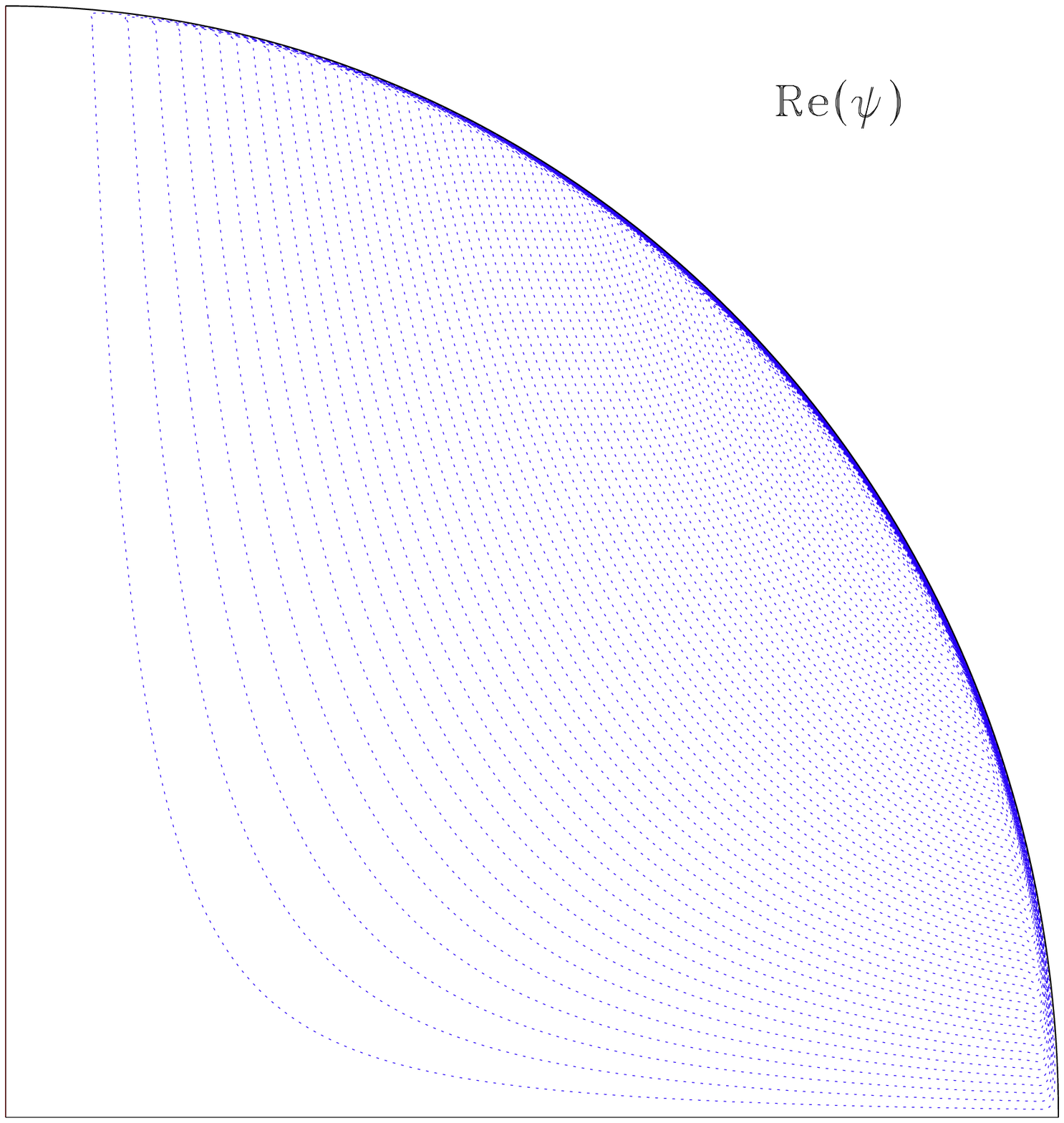}\\
	\end{center}
\caption{Differential rotation $\delta \Omega$ and meridional circulation
stream function $\psi$ (red : direct sense, blue: clockwise sense) shown
in the meridional plane for $E=10^{-6}$ and $b=\{1,-1\}$ (top and bottom
respectively). The stellar rotation axis is vertical.\label{fig2}}
\end{figure}

Solving numerically Eqs. (\ref{eq2}) and (\ref{eq3})
leads to velocity fields that can be described by a
differential rotation $\delta\Omega=u_\varphi/s$
and the stream function $\psi$ of the meridional flow.
As we can see in Fig. \ref{fig2}, the differential rotation is cylindrical as the Taylor-Proudman theorem predicts. The meridional circulation has a unique cell in each hemisphere
whose circulation sign depends also on the sign of $\Delta\Omega$.
Namely, the circulation is clockwise when $\Delta\Omega<0$ and counter
clockwise when $\Delta\Omega>0$.

\begin{figure}
 	\begin{center}
 \includegraphics[scale=0.4]{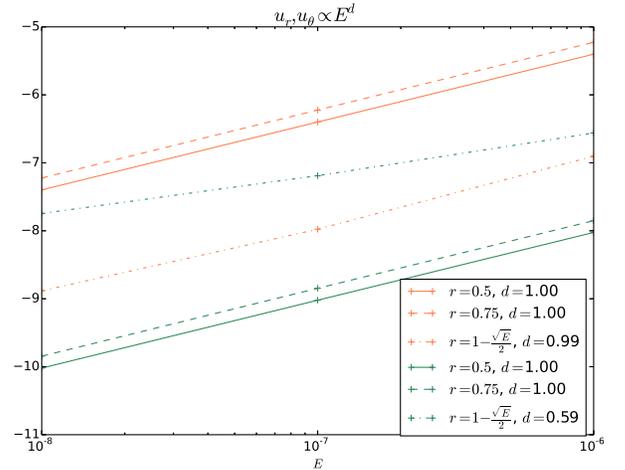}
	\end{center}
\caption{Logarithm of the absolute value of the radial (orange lines) and latitudinal
(green lines) components of the velocity field as a function of $E$
at $r=\{0.5, 0.75,1-\frac{\sqrt{E}}{2}\}$, $\theta=\pi/2$ and for
$b=+1$. Scaling laws are $\{u_r,~u_\theta\} \propto E^d$, where the index $d$
is indicated for each case on the legend panel. \label{fig4} }
\end{figure}

We study the dependence of the velocity field amplitude with the Ekman
number and summarize the obtained results in Fig.~\ref{fig4}.  The amplitude
of the azimuthal velocity does not depend on the Ekman number value
as expected.  The amplitude of the meridional circulation does depend
on the Ekman number value and is proportional to $E$ as also expected.

Inside the Ekman boundary layer that develops at the outer boundary near $r=1-\frac{\sqrt{E}}{2}$,
one can notice the boundary layer corrections.
The radial component is $\mathcal{O}(E)$, while the
latitudinal one is closer to $\mathcal{O}(\sqrt{E})$.

\begin{figure}
 	\begin{center}
 \includegraphics[scale=0.5]{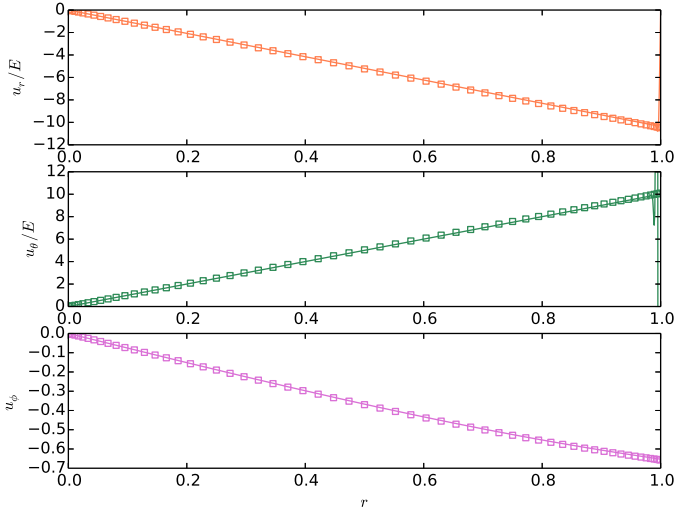}
	\end{center}
\caption{Comparison of the analytical expressions of the velocity
components derived in the last section (squares) and the numerical results
(solid lines) as a function of the radius at the colatitude $\theta=0.5$
for $E=10^{-6}$ and $b=+1$ in the radial (orange), latitudinal (green) and azimuthal (purple) directions. \label{figComp}}
\end{figure}

In Fig. \ref{figComp}, we show the velocity components in the radial
(orange), the latitudinal (green) and the azimuthal (purple) 
directions obtained numerically (solid lines). We overplot the
analytical expressions that we derived in the last section with squares.
The numerical and analytical solutions are in perfect
agreement except near the outer boundary where boundary layer corrections
apply.

\section{The stably stratified case}
\label{sect2}

We now move to the stably stratified case. For that we insert into the
fluid cold sources \citep[following][]{MR06}, hereafter R06, that set a stable density
stratification and trigger a baroclinic flow. We wish to determine the
parameter regime where the baroclinic differential rotation and the
foregoing viscosity driven differential rotation respectively dominate.

\subsection{The flow equations}

To keep the model simple, we
use the Boussinesq approximation. Thus density fluctuations only appear
in the buoyancy term here driven by the effective gravity (i.e. the
combined effect of centrifugal and gravitational accelerations).

Thus, we introduce heat sinks $\delta Q$ that will
perturb the previous flow $\vV_0 = b|\Delta\Omega|R s^3\ephi$. The
heat equation for a steady-state solution thus reads

\begin{equation}
\vv\cdot\vec{\nabla}\delta T = \kappa\Delta\delta T + \delta Q\; ,
\end{equation}
where $\delta T$ is the temperature perturbation generated by the heat
sinks and $\kappa$ is the heat diffusivity of the fluid. In this
equation $\vv = \vV_0+\delta\vv$ where $\delta\vv$ is the velocity
perturbation
arising from the introduction of the heat sinks. Note that
$\delta\vv$ is not necessarily small compared to $\vV_0$. However, we
shall neglect the advection heat term $\vv\cdot\vec{\nabla}\delta T$ altogether
on the ground that we expect that $\delta T$ be axisymmetric and $\vv$
be dominated by its azimuthal component. We already know that the
meridional circulation is \od{E} smaller than the azimuthal flow in the
unstratified case. The baroclinic solutions derived by \cite{MR06}
share the same property, so we can confidently expect that this nonlinear
term is small (see below for the discussion). We are
thus lead to a simple equation for the steady temperature field introduced
by the heat sinks, namely

\begin{equation}
\kappa\Delta\delta T + \delta Q = 0\; ,
\end{equation}
which is solved by $\delta T(r)$, where we assume that the heat sinks
$\delta Q$ have a spherically symmetric distribution. Let us now move to
the equations for the velocity field. Mass conservation still demands

\begin{equation}
 \vec{\nabla}\cdot\vv = 0 \; ,
 \label{eqcontpoura}
\end{equation}
because of the use of the Boussinesq approximation. The momentum
equation now reads

\begin{equation}
\rho(\vv\cdot\vec{\nabla}\vv + 2\vO_0\wedge\vv) = -\vec{\nabla} P +\rho\vg_{\rm eff} +
\mu\Delta\vv \; , \label{mom}
\end{equation}
which is written in a frame rotating at angular velocity $\vO_0$, i.e.
the angular velocity of the pole. We have now included the associated
centrifugal acceleration into the effective gravity $\vg_{\rm eff}$.

The density is perturbed by the temperature variations so that 

\begin{equation}
\rho = \rho_0 +\delta\rho\; .
\end{equation}
Here $\rho_0$ is constant and associated with the reference temperature
$T_0$. So we also use the simple equation of state (usually associated
with the Boussinesq approximation)

\begin{equation}
 \frac{\delta\rho}{\rho_0}= -\alpha(T-T_0) = - \alpha\delta T\; ,
\end{equation}
where $\alpha$ is the dilation coefficient at constant pressure.

As the \BA commands it, we neglect $\delta\rho$ everywhere except
in the buoyancy term. Thus, we rewrite Eq. (\ref{mom}) as

\begin{equation}
\rho_0[(\vv\cdot\vec{\nabla})\vv + 2\vO_0\wedge\vv] = -\vec{\nabla} \Pi +\delta\rho\vg_{\rm eff} +
\mu\Delta\vv \; , \label{momb}
\end{equation}
where $\Pi$ is a reduced pressure which now include the barotropic term
$\rho_0\vg_{\rm eff}$. Finally, Eq. (\ref{momb}) is rewritten

\begin{equation}
(\vv\cdot\vec{\nabla})\vv + 2\vO_0\wedge\vv = -\vec{\nabla} \Pi/\rho_0 - \alpha\delta
T\vg_{\rm eff} + \nu\Delta\vv\; .
\end{equation}
This equation can be further simplified by remarking that $\delta
T\equiv\delta T(r)$ and that $\vg_{\rm eff} = \vg_0(r)+\Omega_0^2s\es$.
The spherically symmetric part of the buoyancy force can be incorporated
in the reduced pressure, so that only the centrifugal force term needs
to be kept. Finally, taking the curl of this equation we eliminate the
reduced pressure gradient and obtain

\begin{equation}
\vec{\nabla}\wedge[(\vv\cdot\vec{\nabla})\vv + 2\vO_0\wedge\vv-\nu\Delta\vv] =
-\alpha\Omega_0^2\delta T'(r)r\sth\cth\ephi \; ,\label{rot_eq}
\end{equation}
where the prime indicates a radial derivative. In this equation the new driving
by the baroclinic torque appears explicitly.

We may now introduce the Brunt-V\"ais\"al\"a frequency, which quantifies
the stratification of the fluid, namely

\begin{equation}
N^2(r)=\alpha \delta T'g(r)\; ,
\label{n2}
\end{equation}
where $g(r)= r g_s$ for a constant density fluid and with $g_s$ the
gravity at the surface of the sphere. Eq. (\ref{rot_eq}) now reads

\begin{equation}
\vec{\nabla}\wedge[(\vv\cdot\vec{\nabla})\vv + 2\vO_0\wedge\vv-\nu\Delta\vv] =
-N^2(r)\frac{\Omega_0^2r}{g(r)}\sth\cth\ephi\; .
\label{rot_eqb}
\end{equation}

\subsection{Scaled equations}
 
As for the unstratified case we now rescale the equations using
$|\Delta\Omega|R$ as the velocity scale and $R$ as the length scale. We
thus find

\begin{equation}\left\{ \begin{array}{l}
\vec{\nabla}\wedge[\mathcal{R}o\,(\vu\cdot\vec{\nabla})\vu + \ez\wedge\vu - E\Delta\vu] =\\
\hspace*{4cm}  -\varepsilon n^2(r) \sth\cth\ephi \; ,\\
\\
\vec{\nabla}\cdot\vu = 0 \; ,\label{rot_eqc}
\end{array}\right. 
\end{equation}
where we introduced the scaled \BVF

\begin{equation}
n^2(r) = \frac{N^2(r)}{2\Omega_0|\Delta\Omega|}=\frac{N^2(r)}{4\Omega_0^2}\mathcal{R}o^{-1}\; ,
\end{equation}
and the relative amplitude of the centrifugal force, namely

\begin{equation}
\varepsilon = \frac{\Omega_0^2R}{g_s}\; .
\end{equation}
In the Sun, $\varepsilon_\odot\sim 10^{-5}$.
If, as in section 2, we consider only the limit of small Rossby numbers,
we can linearise system (\ref{rot_eqc}) and solve

\begin{equation}\left\{ \begin{array}{l}
\vec{\nabla}\wedge(\ez\wedge\vu - E\Delta\vu) =
-\varepsilon n^2(r) \sth\cth\ephi\; ,\\
\\
\vec{\nabla}\cdot\vu = 0 \; , \label{rot_eqd}
\end{array}\right. 
\end{equation}
completed with boundary conditions (\ref{scaledbcpsi}) and (\ref{bcu}).
Hence, we assume $\mathcal{R}o\ll1$, but also $\varepsilon n^2\infapp1$,
which is actually possible (see below).

At this stage we may remark that unlike in the case treated in
\cite{MR06}, the heat equation has disappeared and the Prandtl
number does not appear in the problem. The reason is that we neglected
at the outset the heat advection by meridional currents. As shown in
\cite{MR06}, this is strictly valid in the limit of the vanishing
$\lambda$-parameter

\[ \lambda = \PR\frac{N^2}{4\Omega_0^2}\; ,\]
where $\PR=\nu/\kappa$ is the Prandtl number defined as the ratio of the kinematic viscosity to the thermal diffusivity. 
In the Sun, $\lambda_\odot\sim 10^{-2}$.
Moreover, $\lambda\ll1$ is obtained
for stars rotating sufficiently rapidly.
However, if the star is a slow rotator, steady state flows are not
relevant since we know that initial conditions then also control the
actual flow, because baroclinic modes are damped on the Eddington-Sweet
time which tends to infinity as rotation vanishes
\citep[e.g.][]{busse81,MR06b}.

This setting generates a thermal wind solution arising from the baroclinic torque and the geostrophic solution generated by the shear imposed by the boundary conditions described in the previous section. It allows us to evaluate their competition.

\subsection{Stars matching the model}

The foregoing model is rather simple but uses a number of hypothesis. We
now need to identify stars that match these conditions.

\begin{figure*}
	\begin{center}
\includegraphics[scale=0.5]{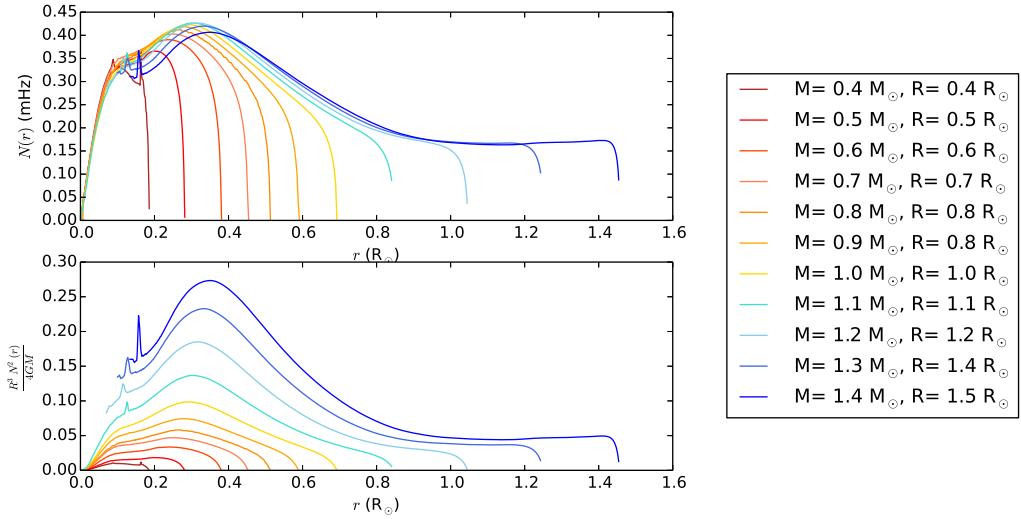}
	\end{center}
\caption{ \textit{Top:} Brunt-V\"ais\"al\"a frequency as a function of the radius, in the radiative core of low-mass stars during the main-sequence ($X_c=0.5$) for stellar masses in the range $[0.4-1.4]M_\odot$ from MESA models. \textit{Bottom:} Amplitude of the right-hand-side of Eq. (\ref{rot_eqd}) (to be multiplied by $\mathcal{R}o^{-1}$) as a function of the radius.\label{AmpRHS}}
\end{figure*}
The right hand side amplitude of Eq. (\ref{rot_eqd}) can be written
\begin{equation}
\varepsilon n^2(r)= \frac{1}{\mathcal{R}o}\frac{R^3N^2(r)}{4GM}\; ,
\end{equation}
where $M$ is the stellar mass and $G$ is the gravitational constant.
We compute main-sequence MESA models \citep{paxton10} with a metallicity of $Z=0.02$ and the mixing-length parameter $\alpha_{MLT}=2$ (the default value used by MESA) when the central hydrogen fraction reaches $0.5$ for masses in the range $[0.4-1.4]M_\odot$, constituting $F$, $G$, $K$ and $M$ stars.
As shown in Fig. \ref{AmpRHS}, the amplitude of the term $R^3N^2/4GM $ in the right hand side
of Eq. (\ref{rot_eqd}) is always less than unity. With small Rossby numbers, the term
$\varepsilon n^2(r)$ can be of order unity, which means that the baroclinic flow has an
amplitude of the same order of magnitude as the geostrophic one driven by the shear.
For young stars, such as ZAMS stars, the amplitude of the term $R^3N^2/4GM$ is of the same order of magnitude leading to the same competition between the baroclinic and the geostrophic flows.
The dynamics set by the shear may be highly modified by the baroclinicity.
For these reason, we need to resort to numerical solutions to determine which flow dominates,
and the corresponding parameter regime. We also determine the features of the dynamics in each identified regime.

\section{Numerical solutions}
\label{sect21}

\subsection{Differential rotation and meridional circulation}

Using the spectral numerical method described in Appendix \ref{MN2}, we numerically solve Eqs. (\ref{rot_eqd}).
We use the $1M_\odot$ MESA model as an input for the right-hand-side of the vorticity equation and we vary the Rossby number systematically.

\begin{figure*}
 	\begin{center}
  \includegraphics[scale=0.17]{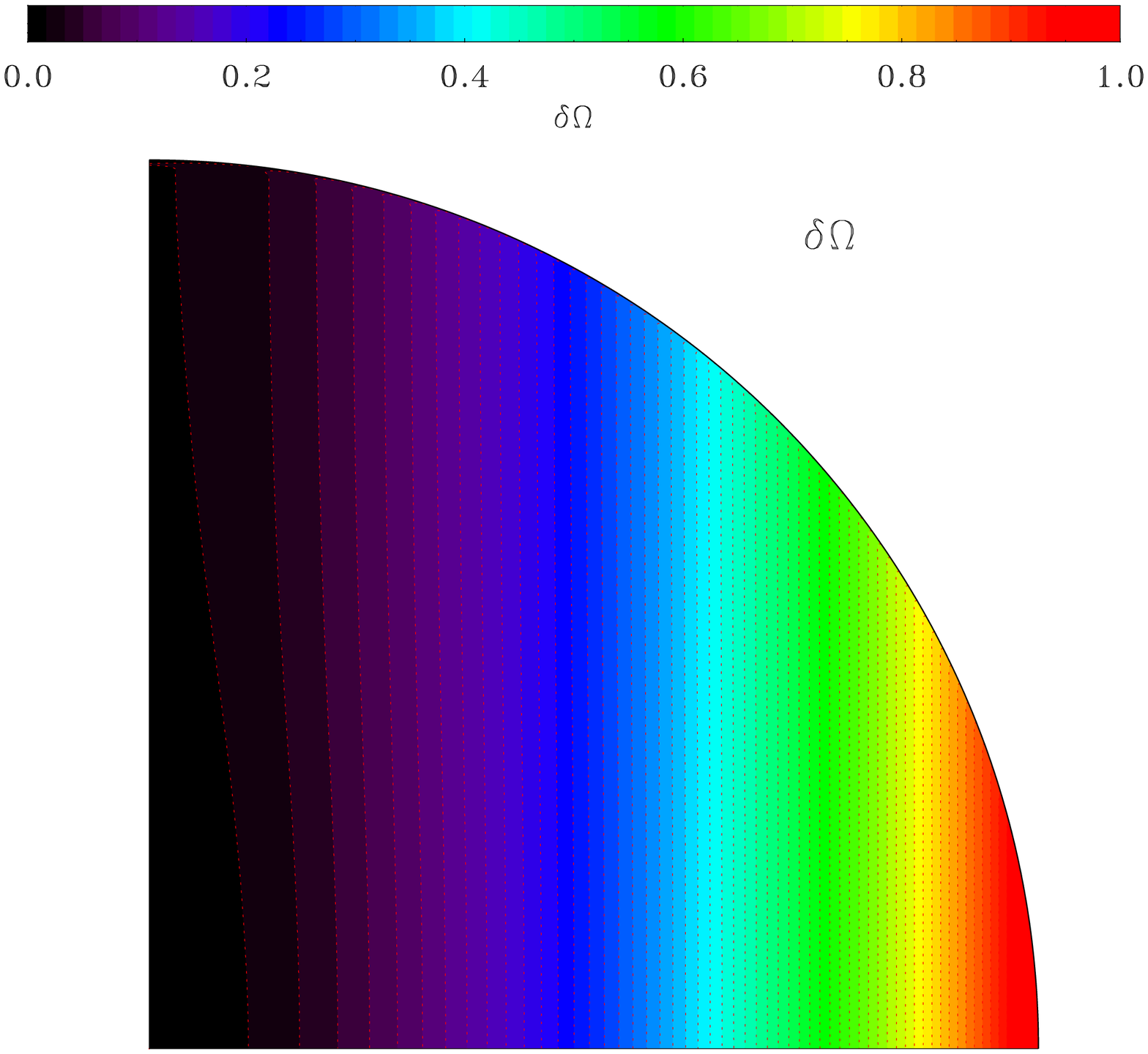}
  \includegraphics[scale=0.17]{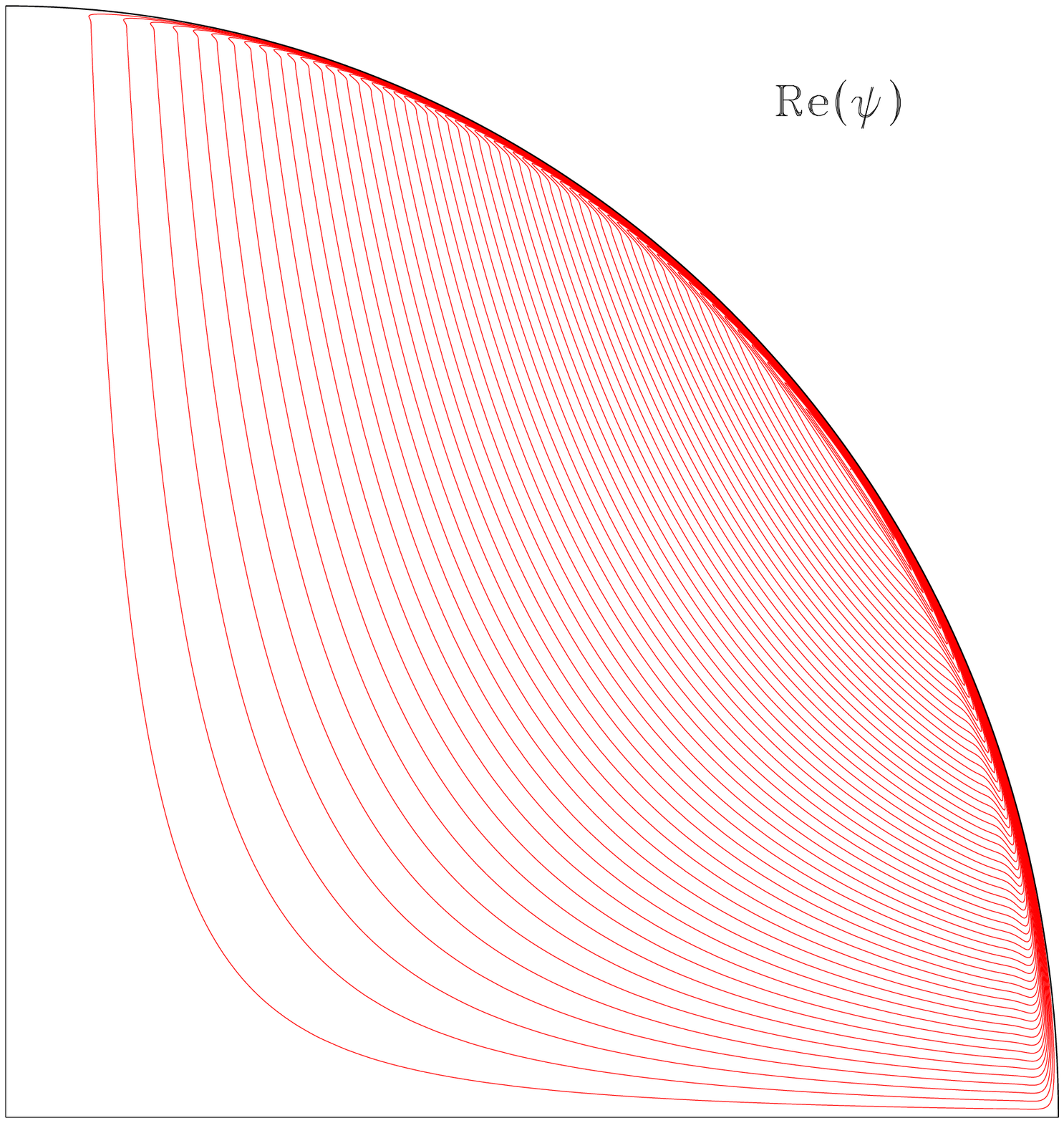}
  \includegraphics[scale=0.17]{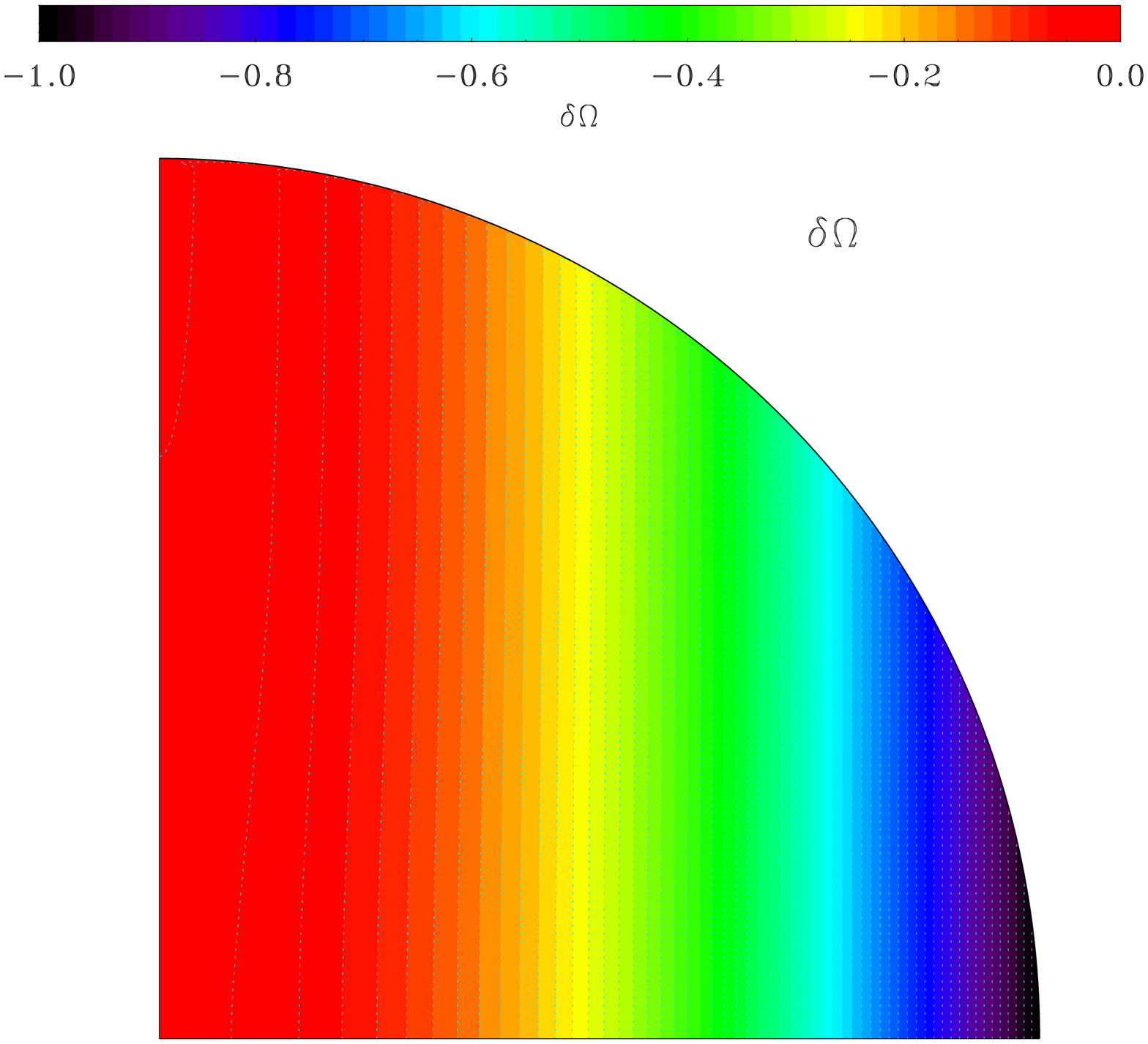}
  \includegraphics[scale=0.17]{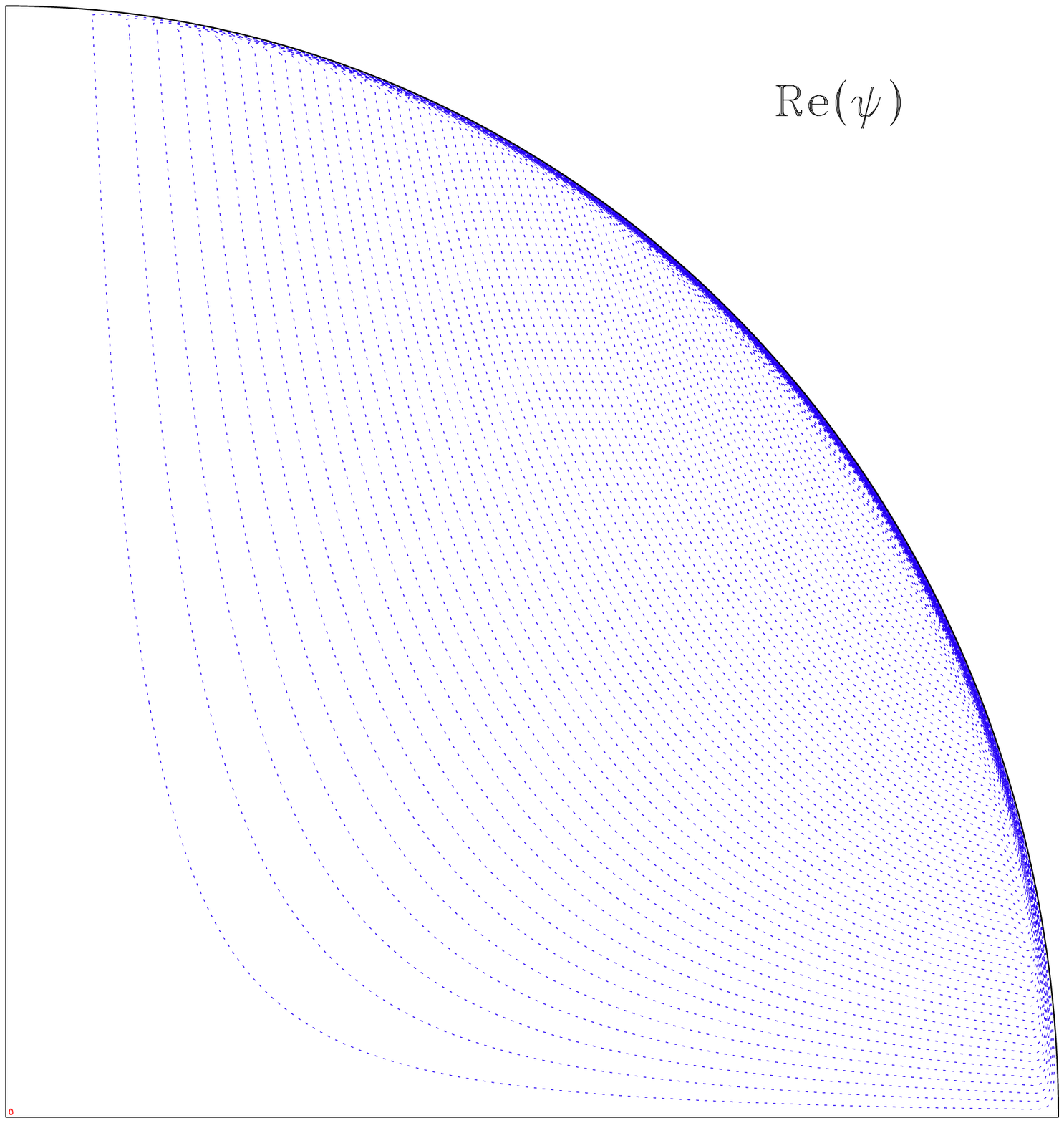}\\
  \includegraphics[scale=0.17]{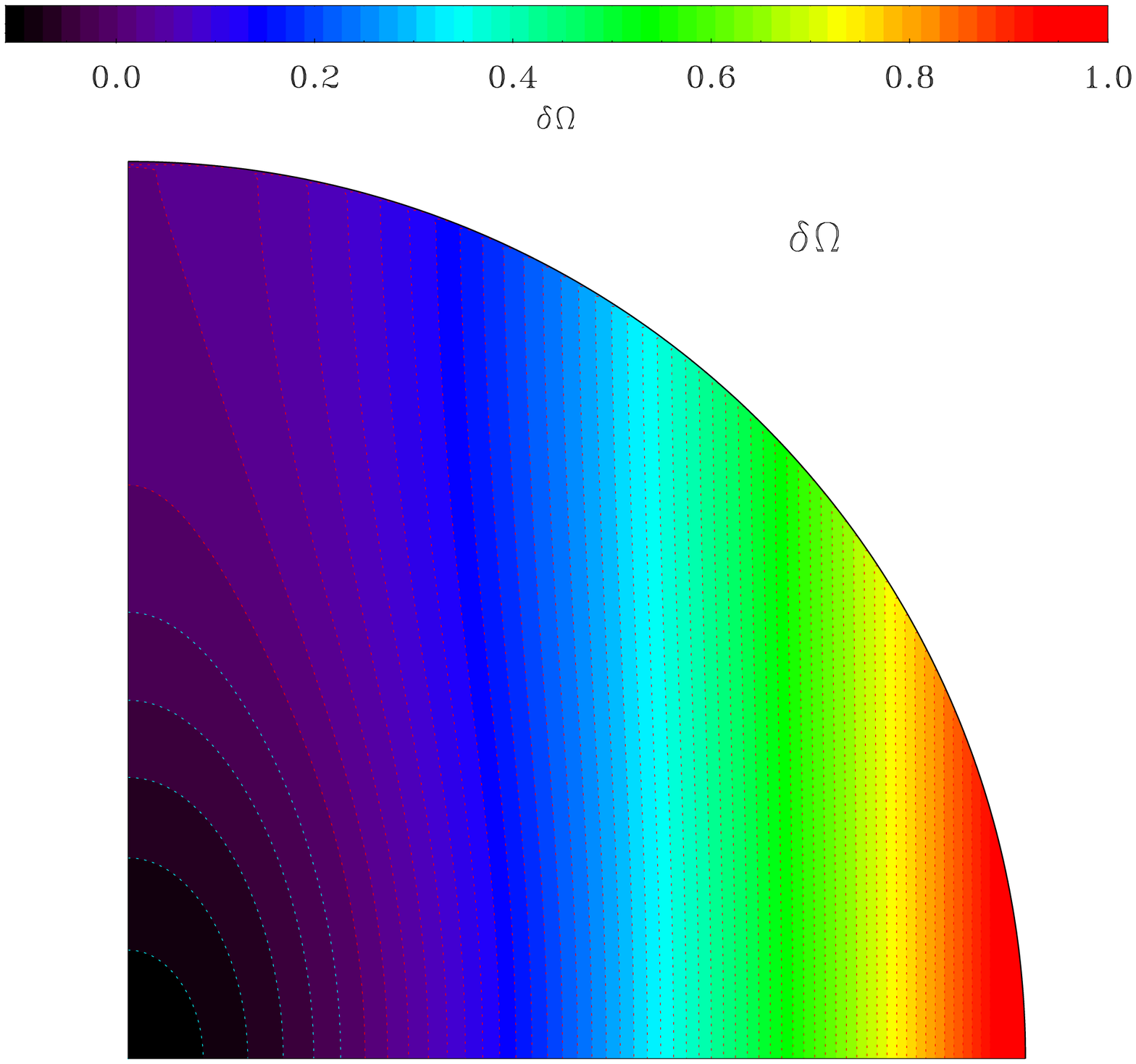}
  \includegraphics[scale=0.17]{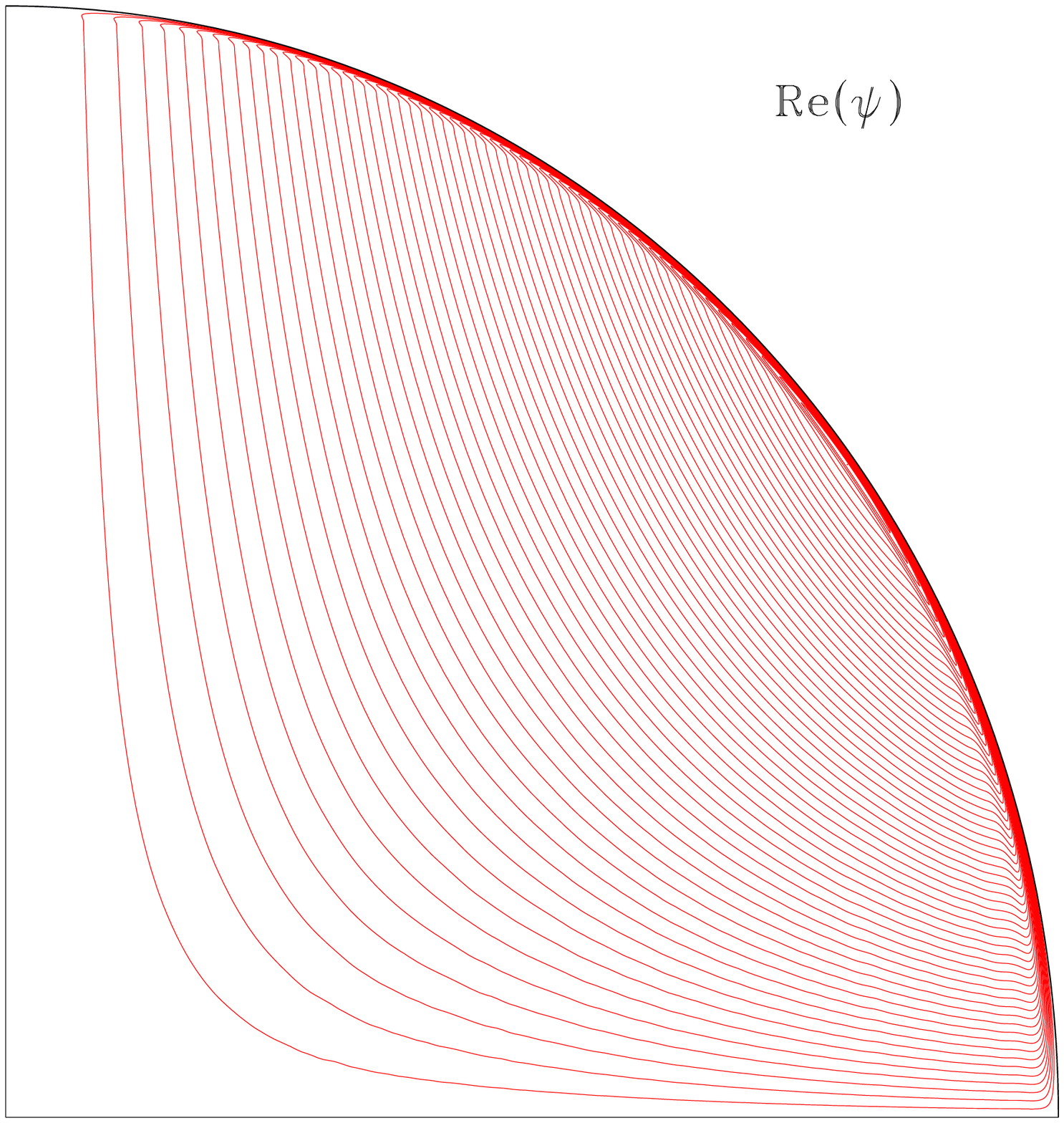}
  \includegraphics[scale=0.17]{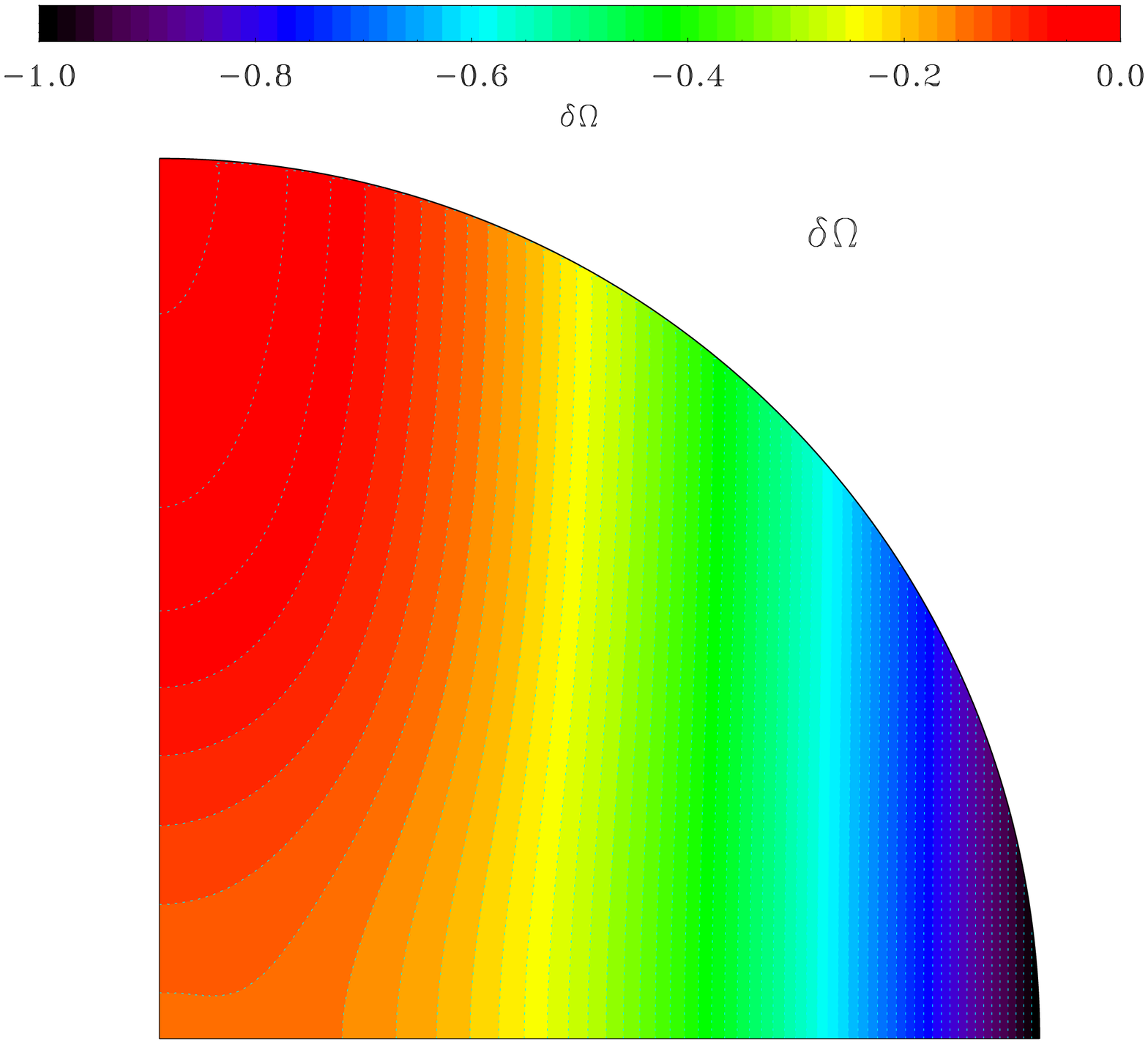}
  \includegraphics[scale=0.17]{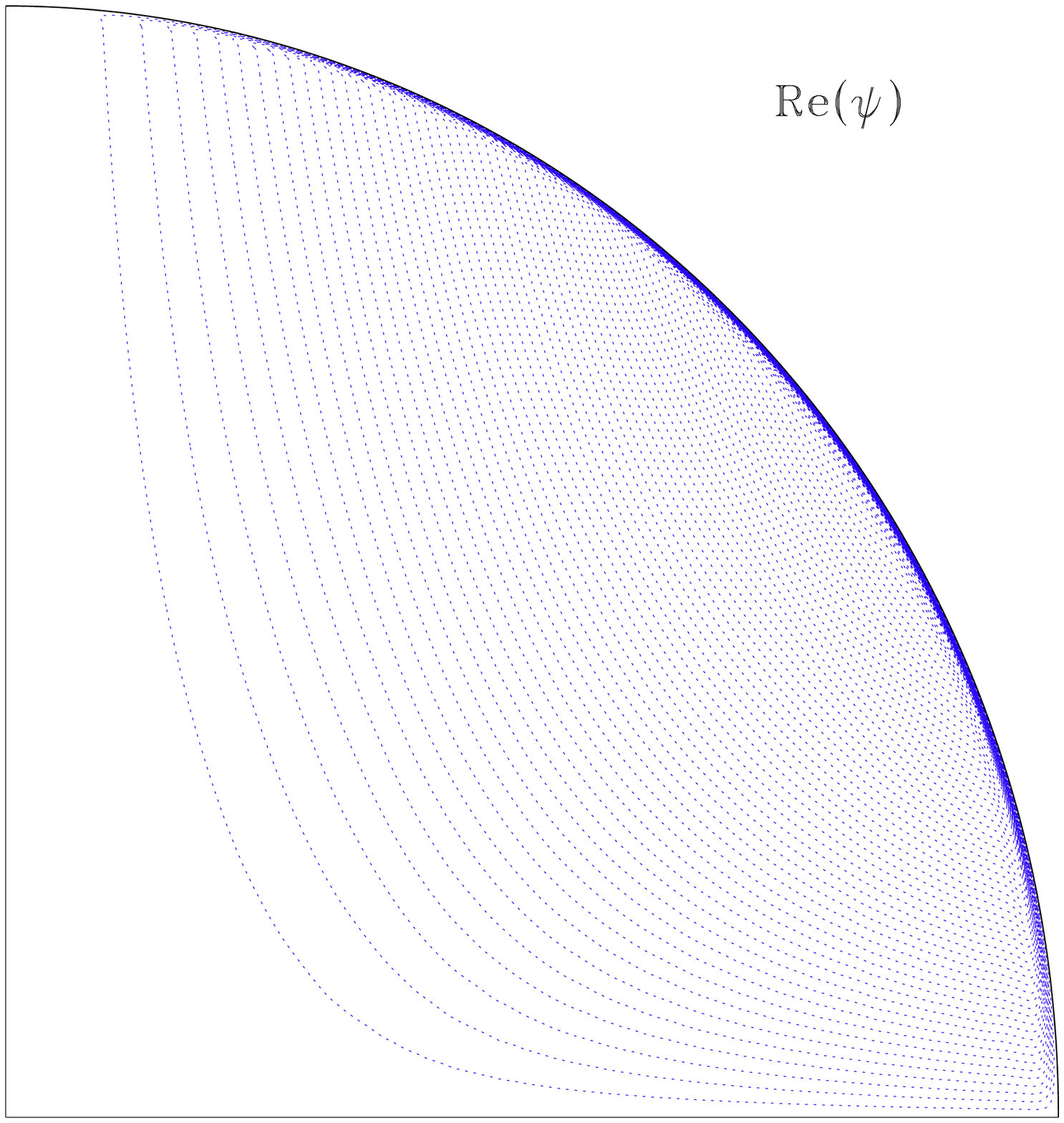}\\
  \includegraphics[scale=0.17]{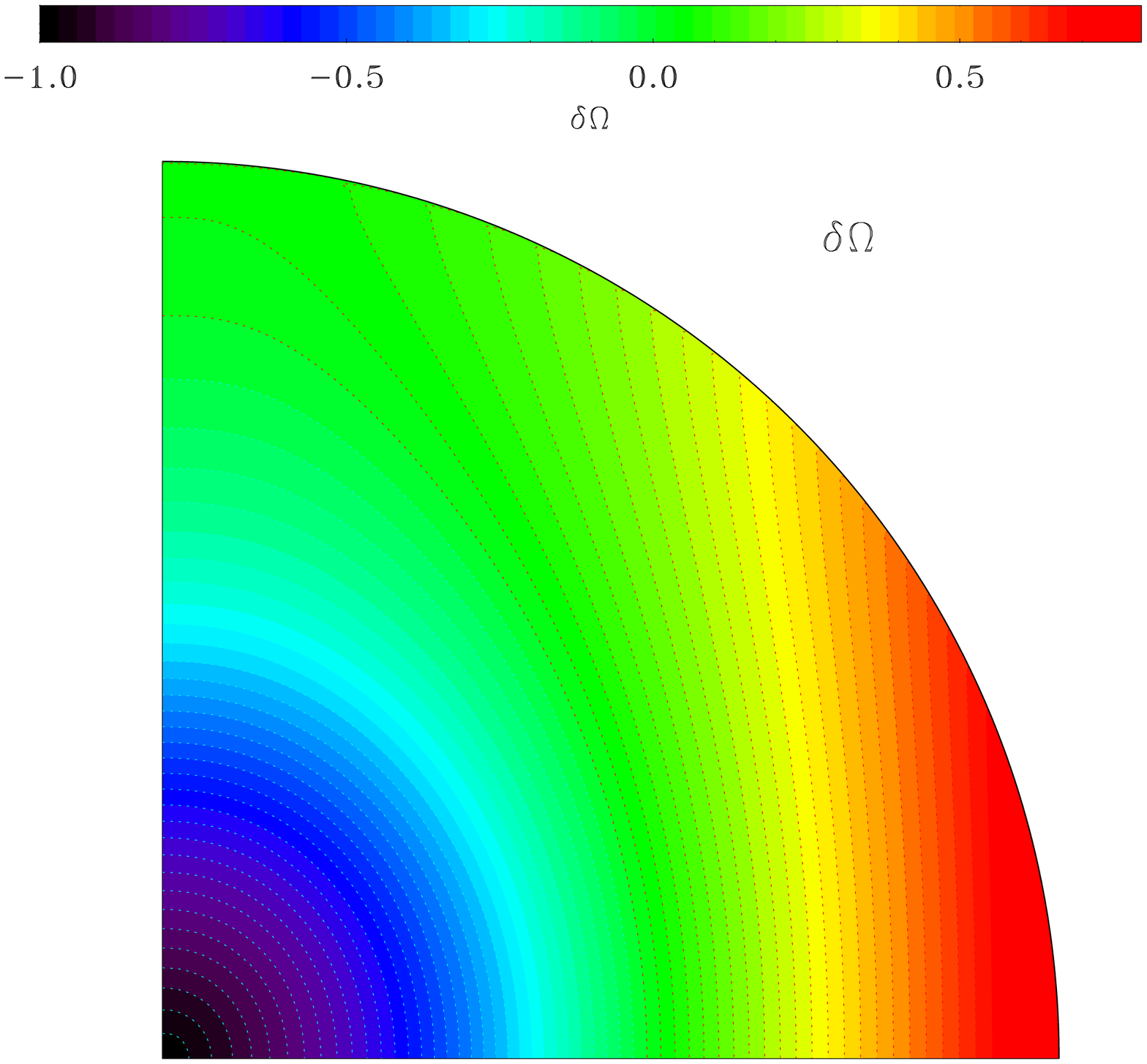}
  \includegraphics[scale=0.17]{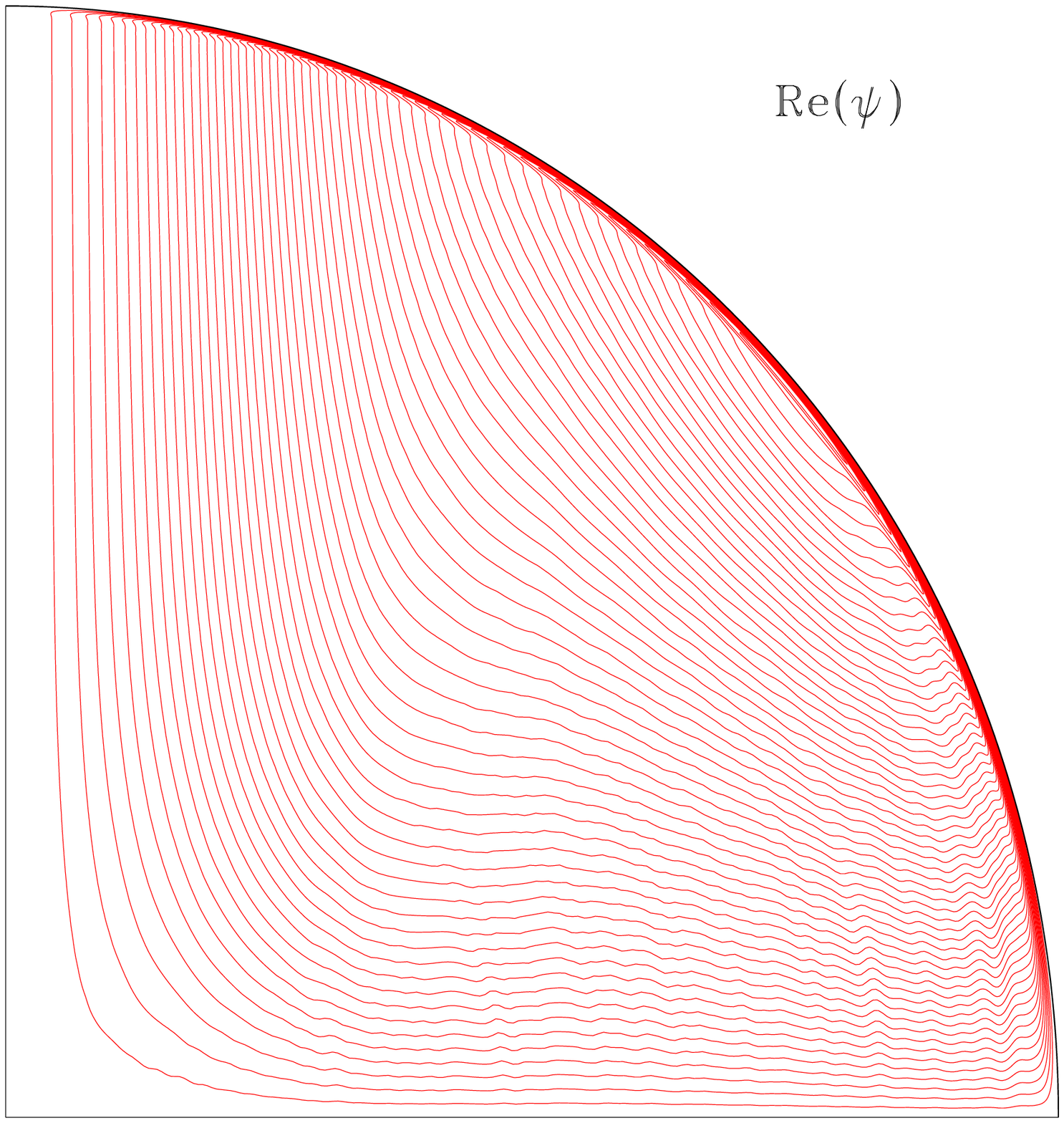}
  \includegraphics[scale=0.17]{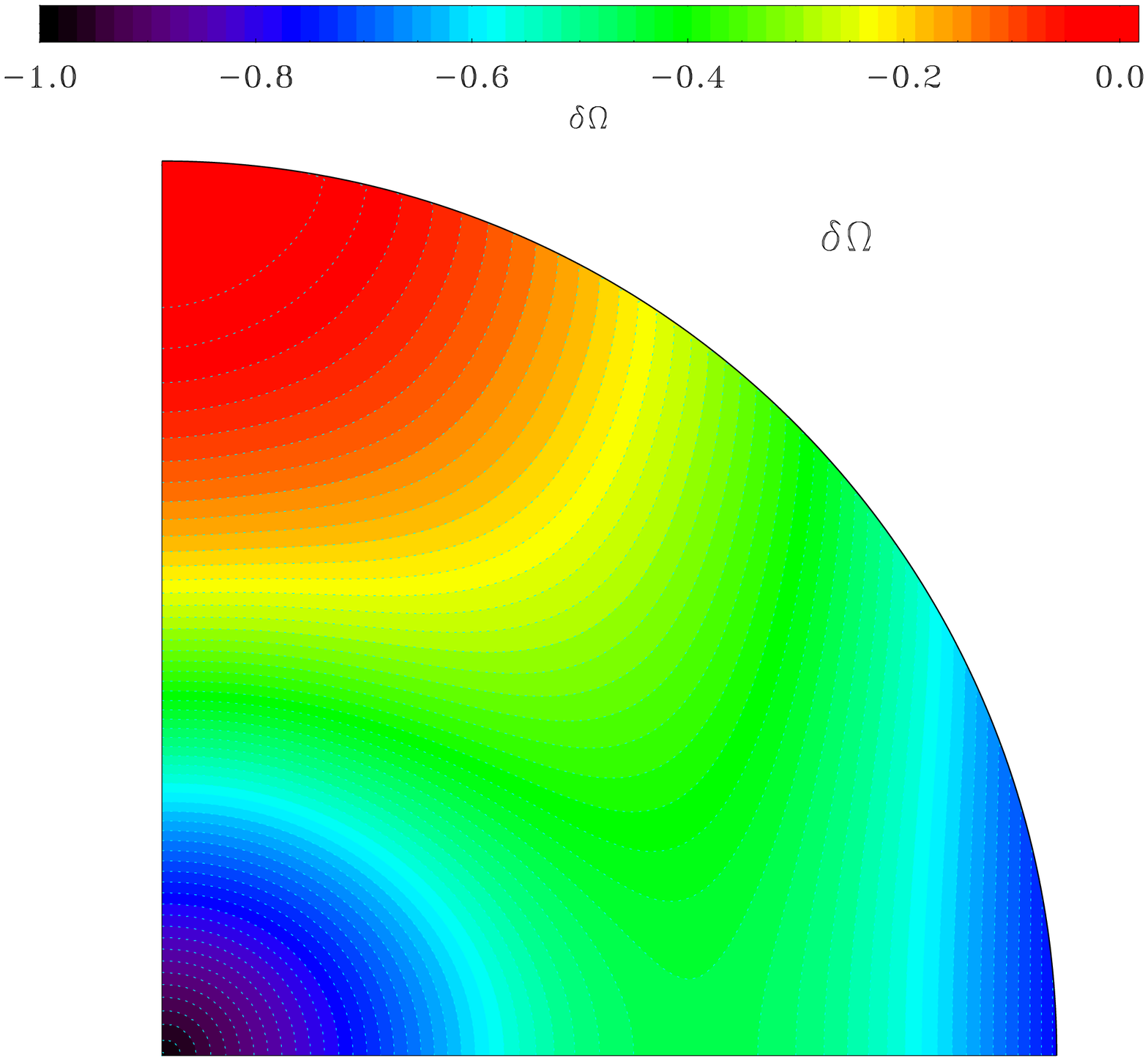}
  \includegraphics[scale=0.17]{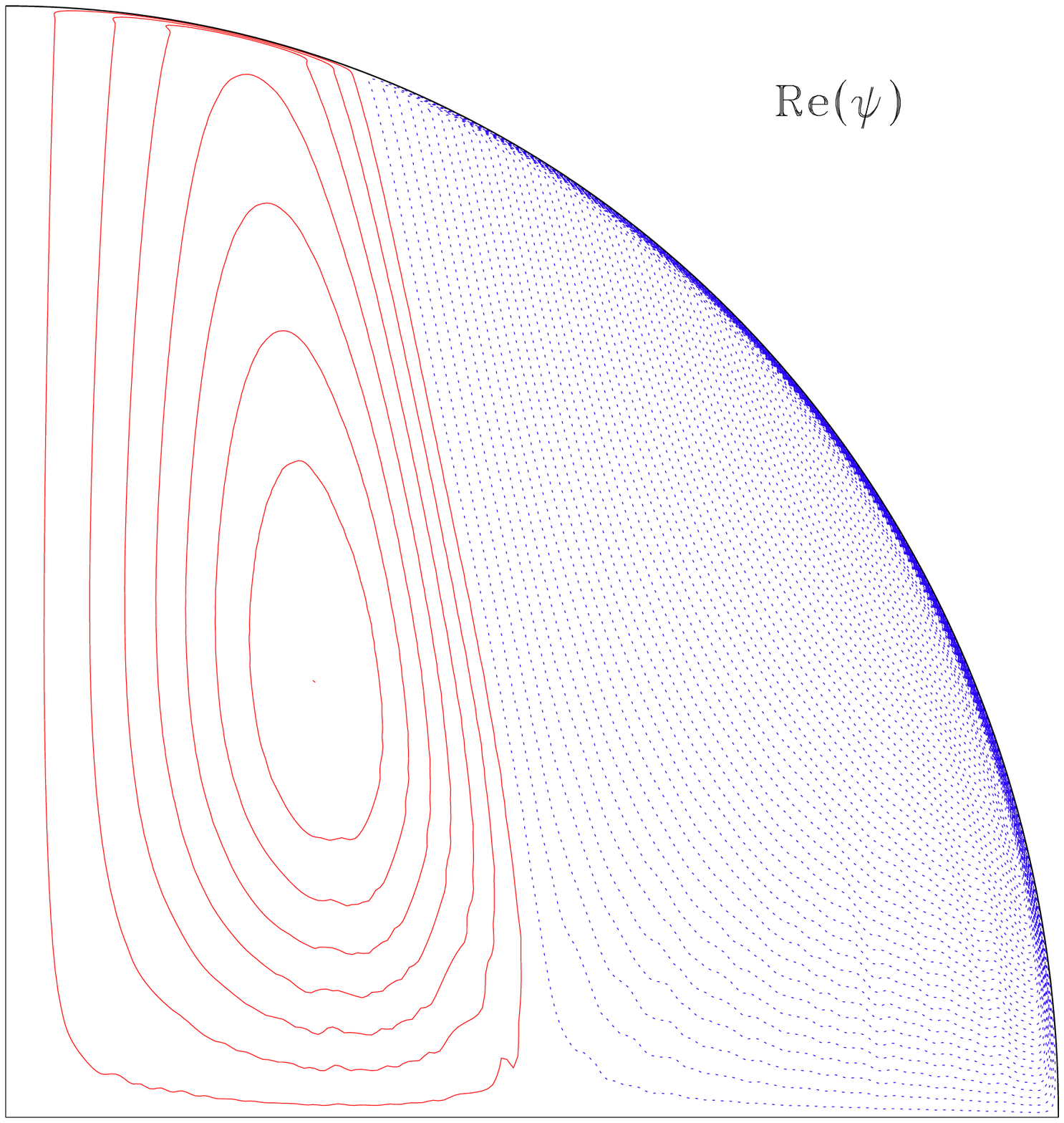}\\         	
  \includegraphics[scale=0.17]{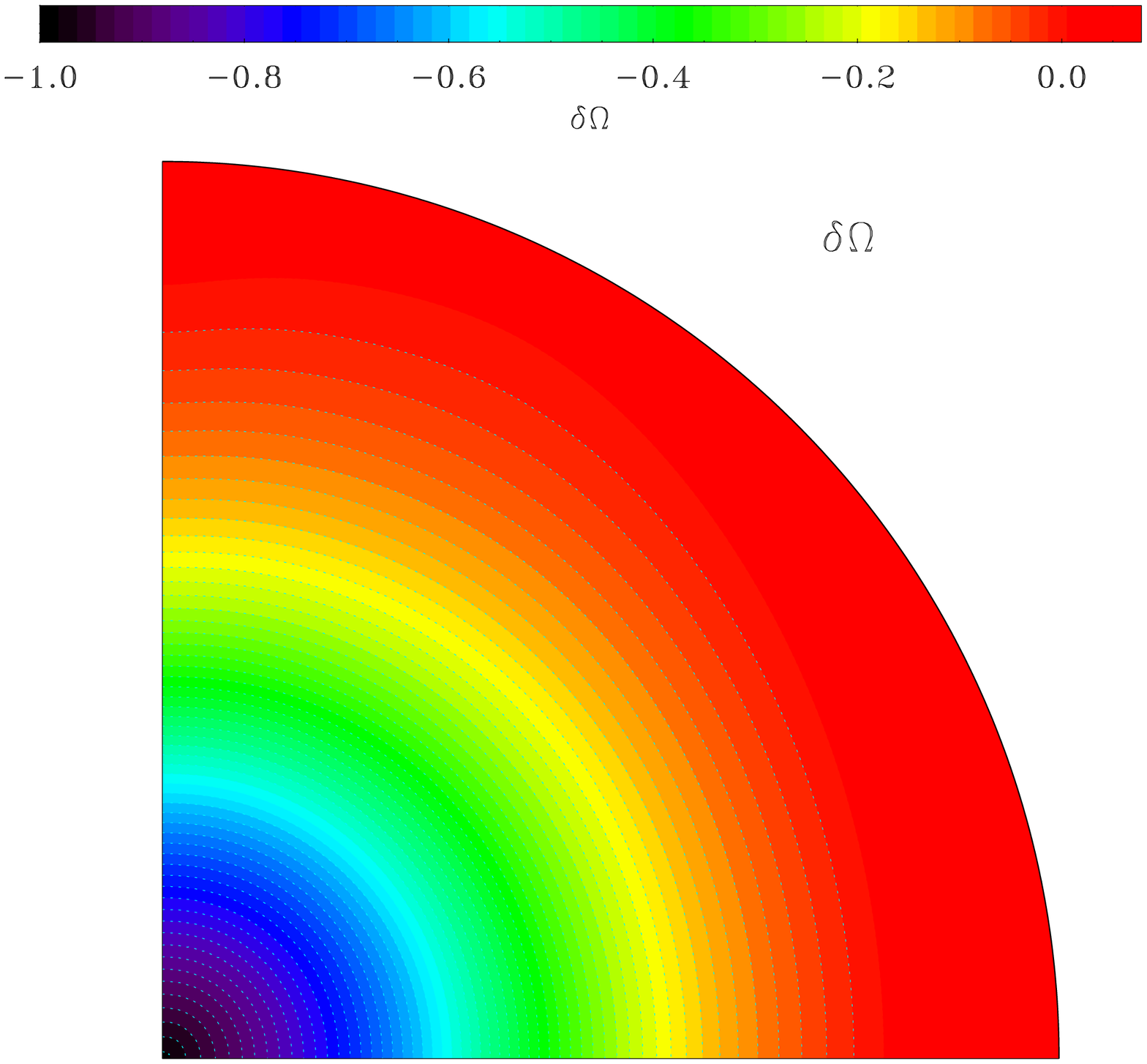}
  \includegraphics[scale=0.17]{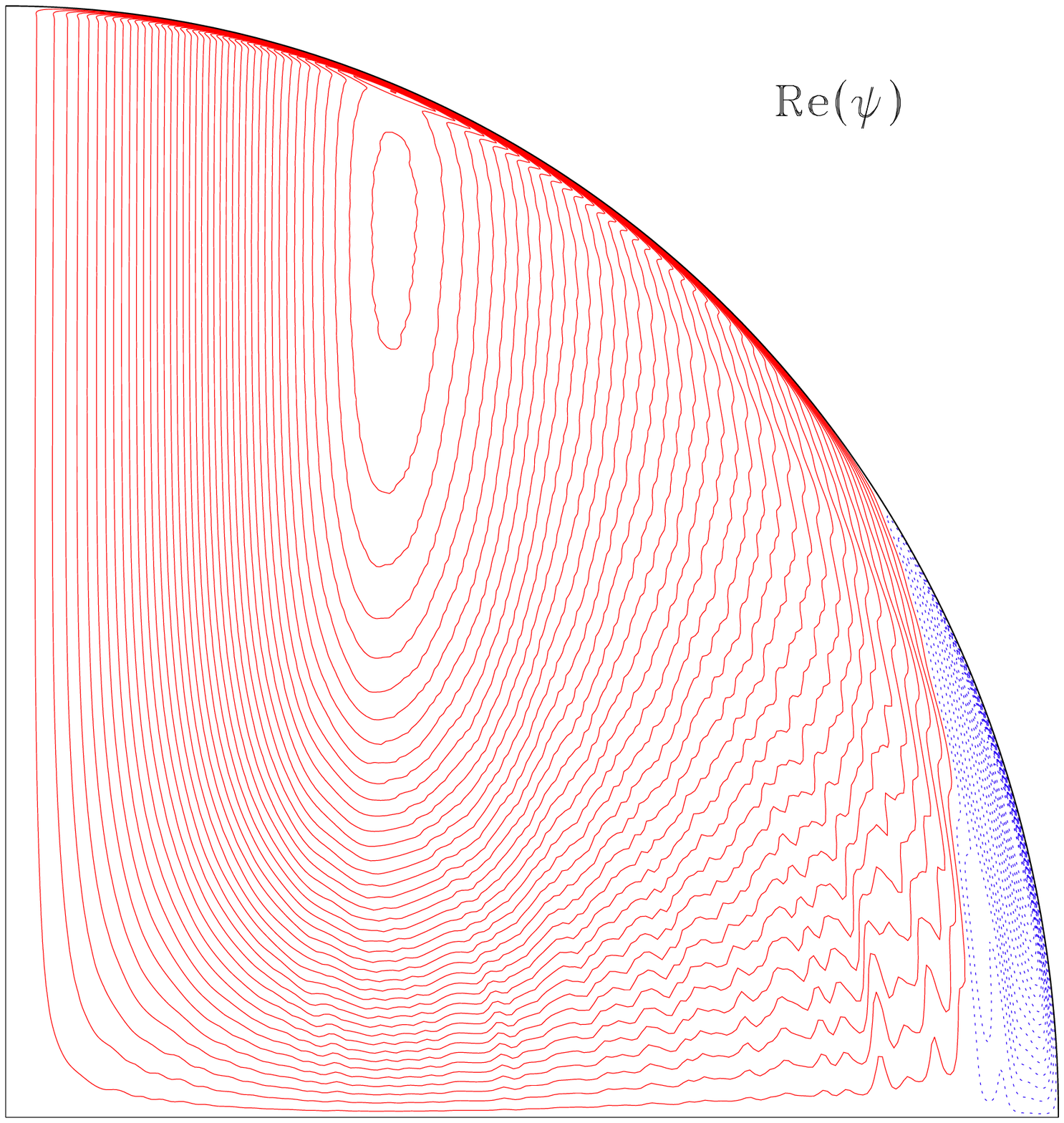}
  \includegraphics[scale=0.17]{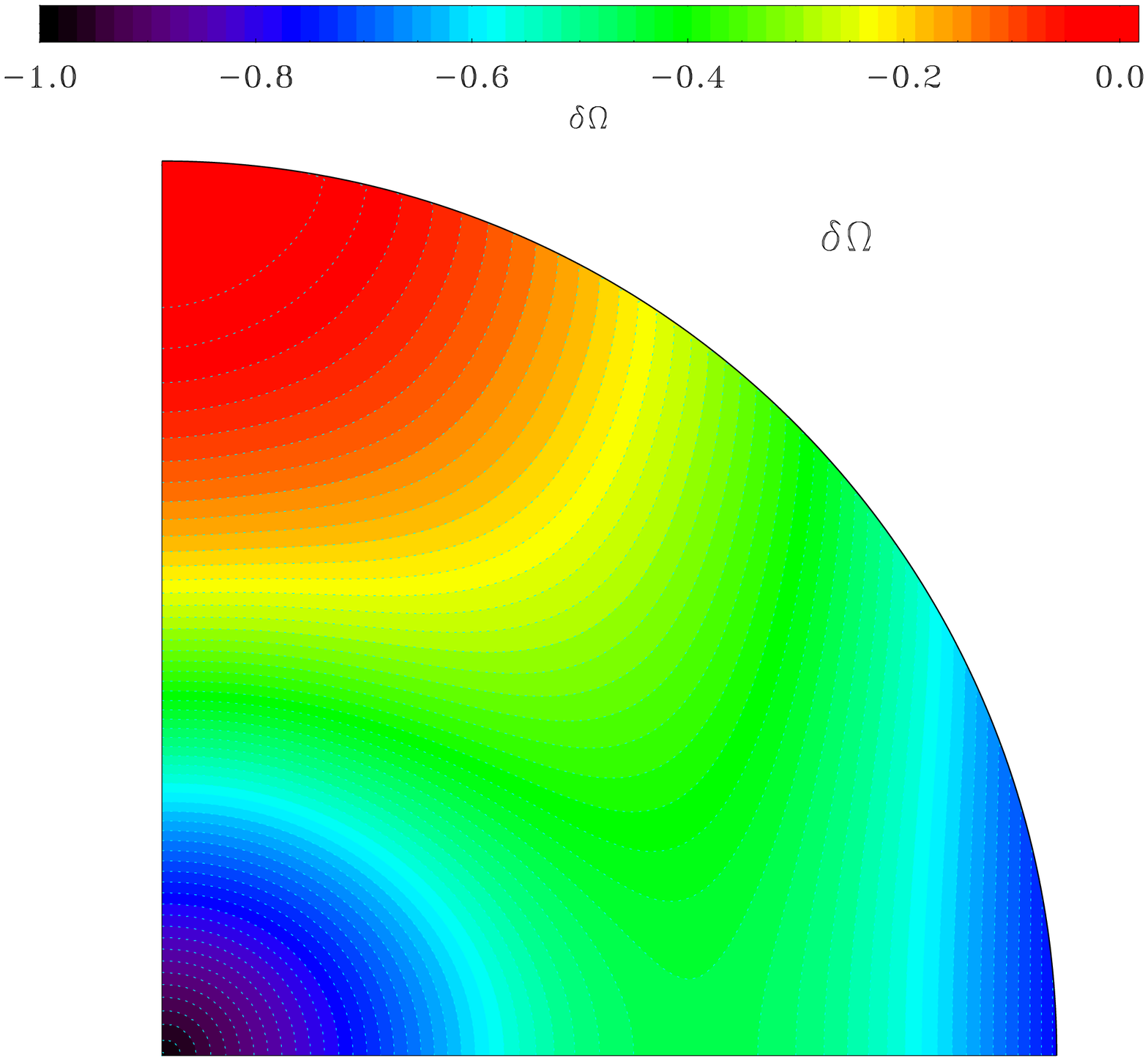}
  \includegraphics[scale=0.17]{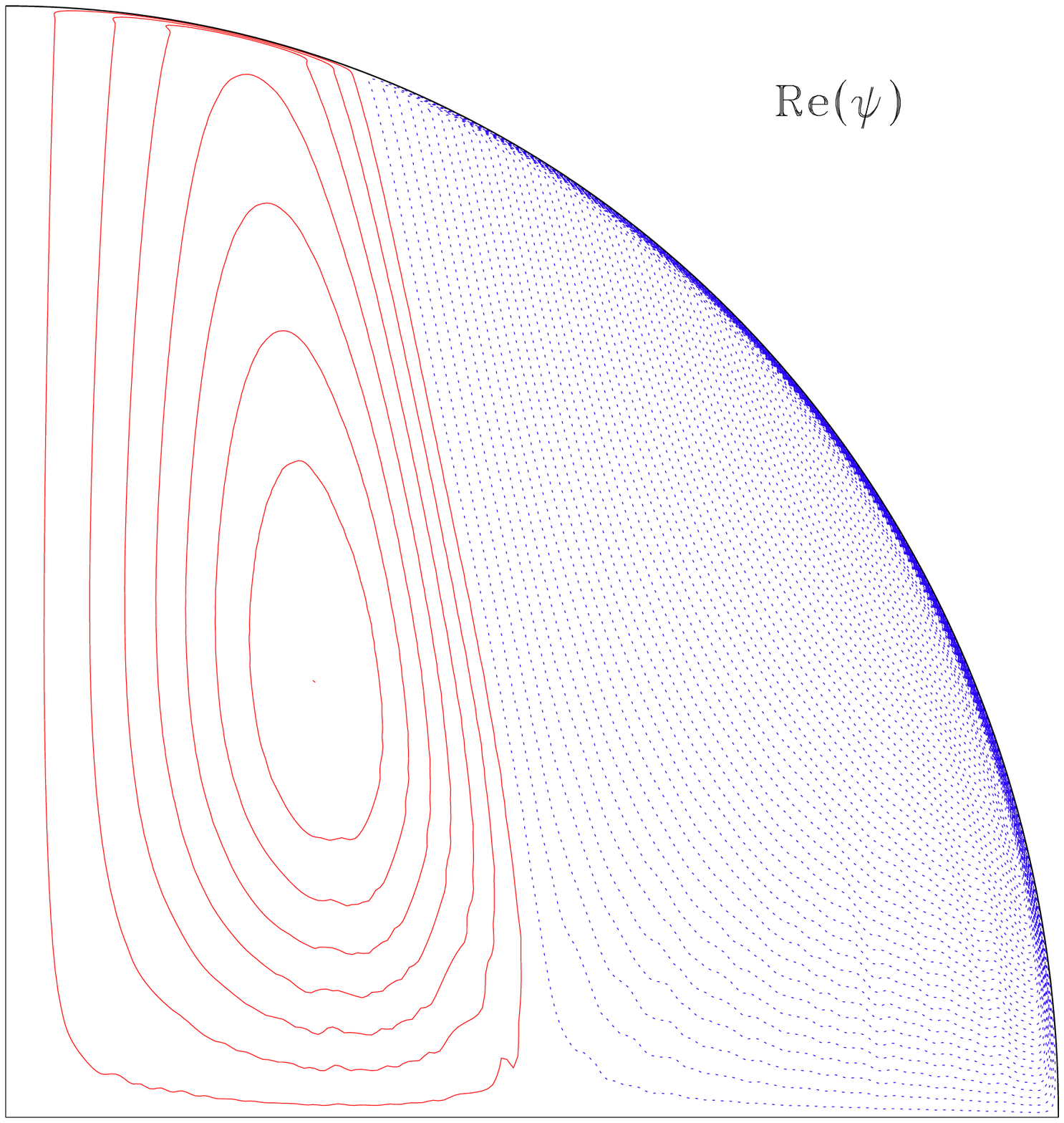}\\
  \includegraphics[scale=0.17]{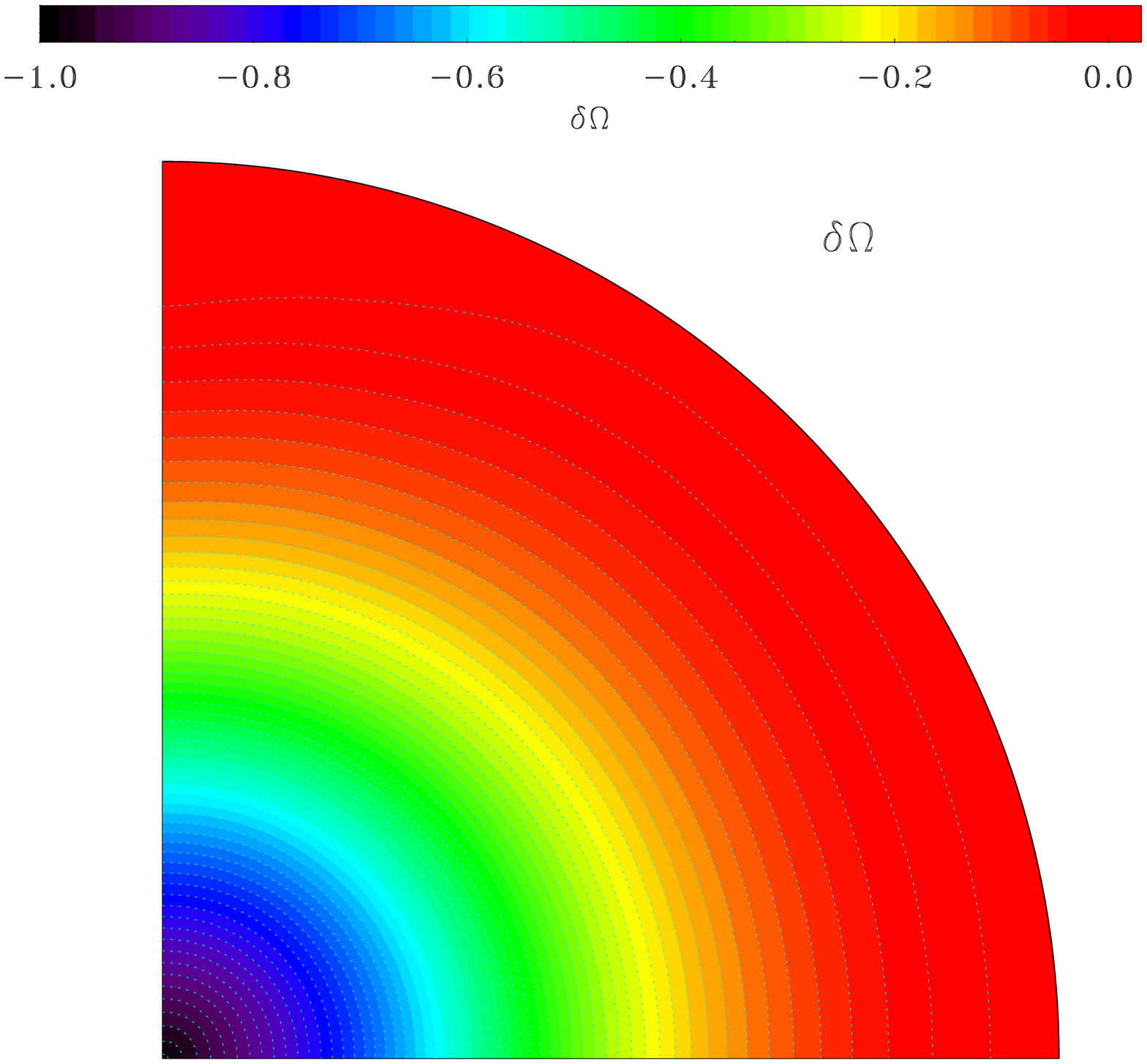}
  \includegraphics[scale=0.17]{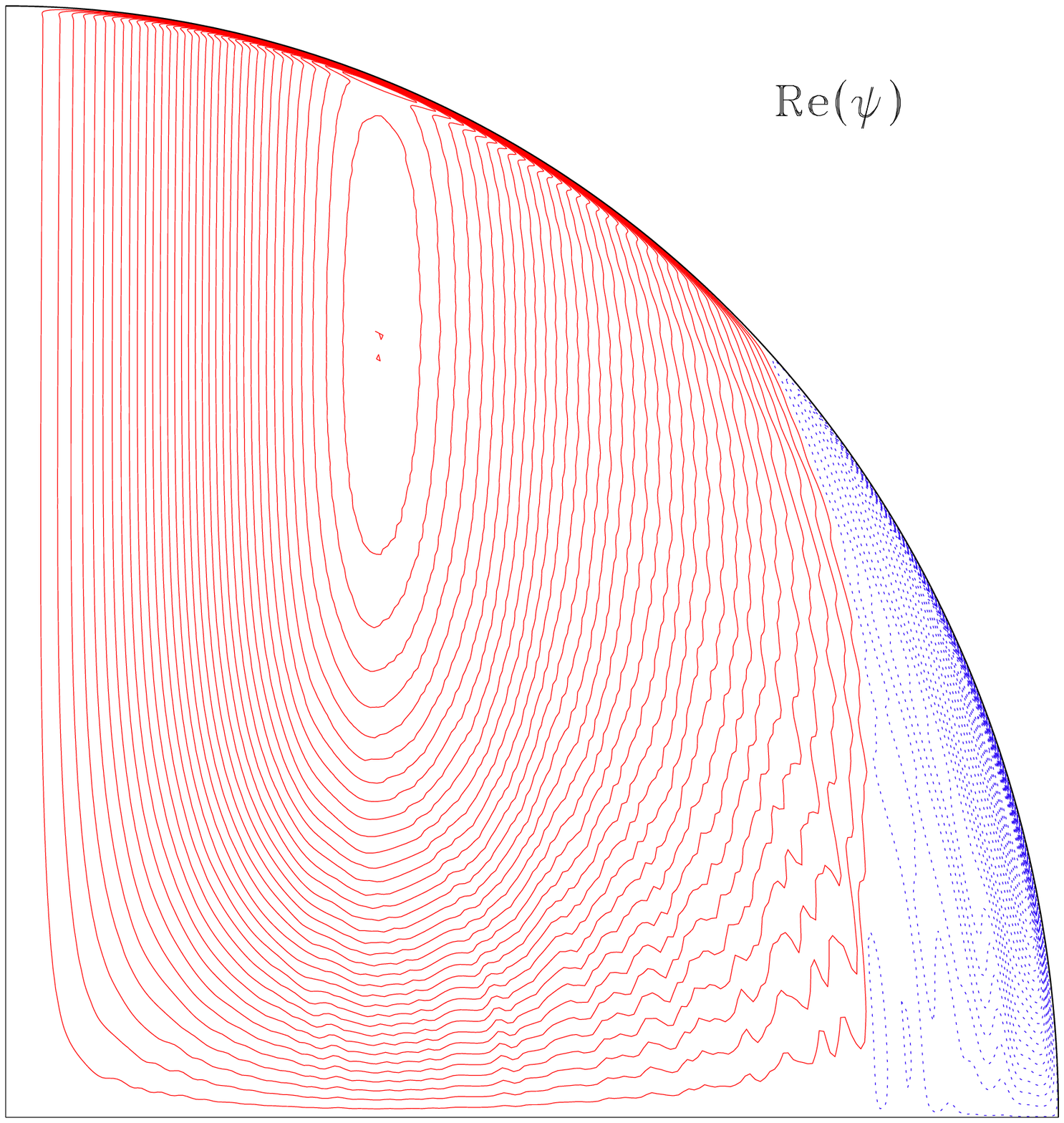}
  \includegraphics[scale=0.17]{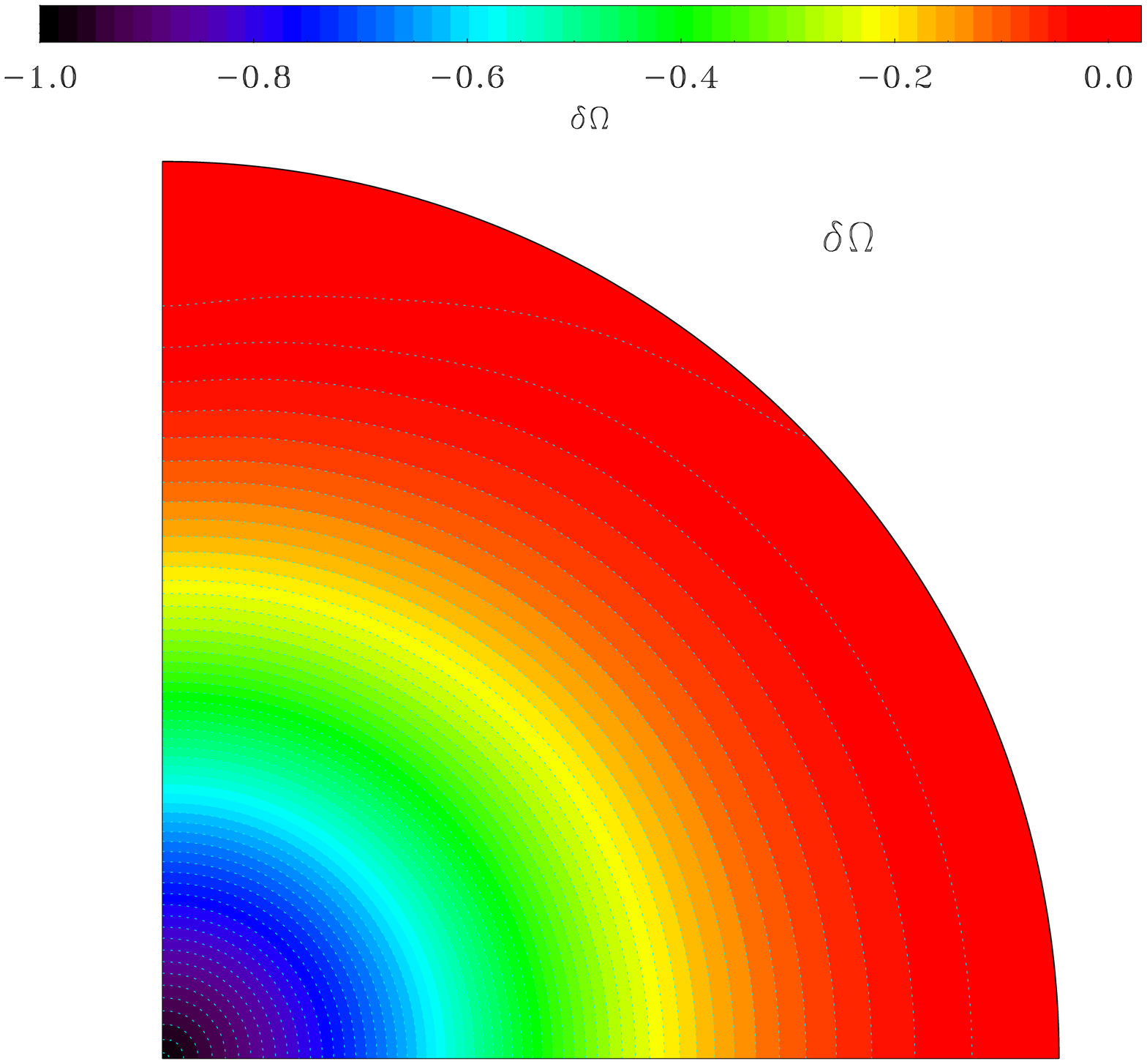}
  \includegraphics[scale=0.17]{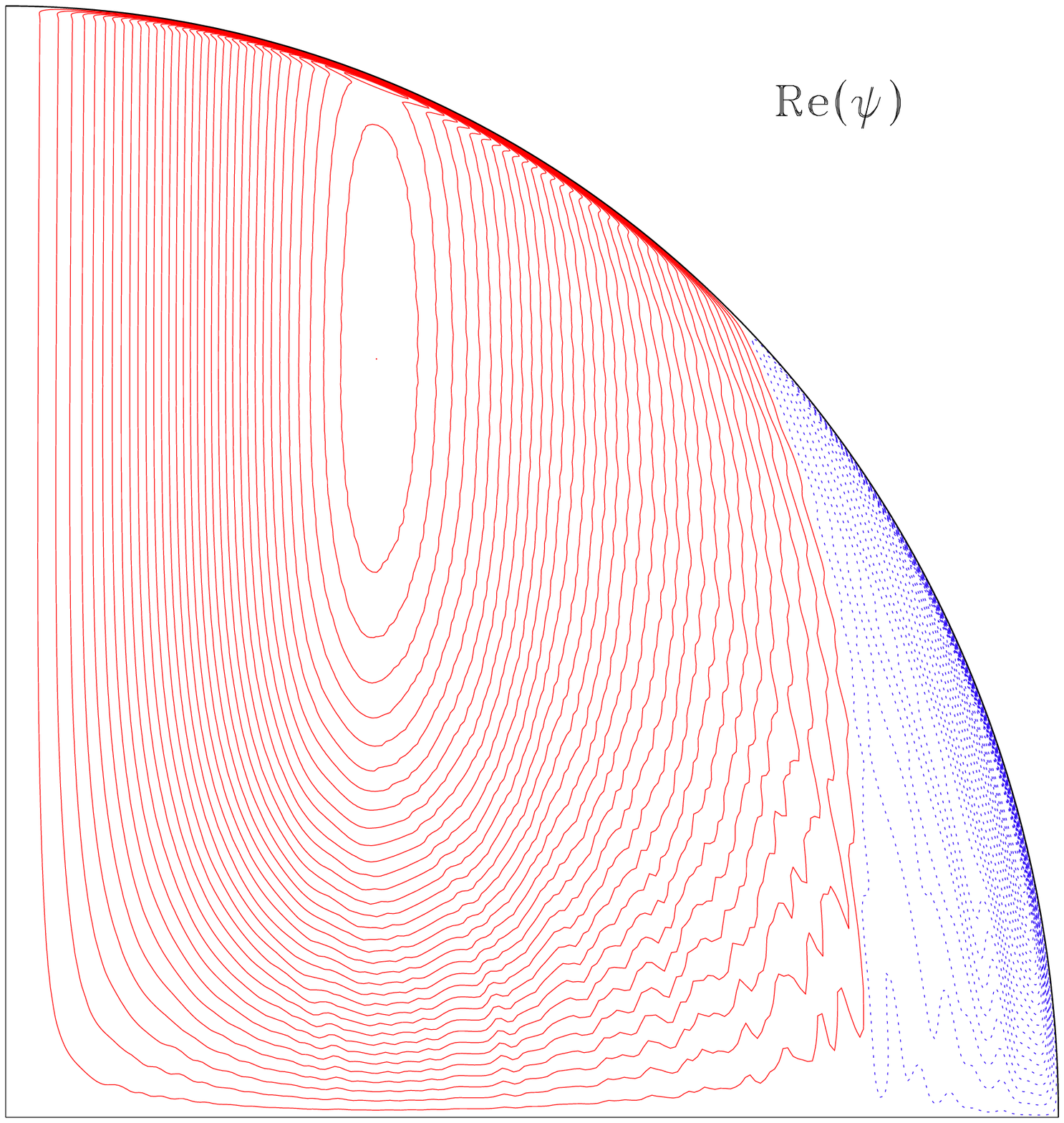}\\
	\end{center}
\caption{Differential rotation $\delta \Omega$ and meridional circulation
stream function $\psi$ (red : direct sense, blue: clockwise sense) shown
in the meridional plane for $E=2.10^{-6}$ and $\mathcal{R}o=\{10, 1, 10^{-1}, 10^{-2}, 10^{-3}\}$ (top to bottom). The two left columns are for $b=+1$ and the two on the right for $b=-1$. The stellar rotation axis is vertical.\label{fig5}}
\end{figure*}

We show in Fig. \ref{fig5}, the differential rotation and the associated meridional circulation for Rossby numbers between $10^{-3}$ and $10$. 
When $\mathcal{R}o\leq 10^{-2}$ (i.e. for weak imposed differential rotation), the dynamics is typical of the flow that arises when a baroclinic torque is applied as described by R06. The thermal wind balance breaks the Taylor-Proudman one.
The differential rotation is roughly shellular and the number of cells of the meridional circulation is equal to the number of inflection points of the Brunt-V\"ais\"al\"a frequency profile plus one, here two. 
These patterns are aligned with the cylindrical z-direction because of rotation.

Differences with the Fig. 4 of R06 come from the upper boundary conditions we set. As shown by Eq. (\ref{scaledbcpsi}), we impose a surface shear on the azimuthal velocity coupled with no penetrative boundary conditions on the meridional components of the velocity field while R06 imposes regular stress-free conditions. 

For Rossby numbers higher than the threshold $\mathcal{R}o\approx 1$ (i.e. for strong imposed differential rotation), the differential rotation profile tends to be cylindrical. The thermal wind from the baroclinic torque is weaker in comparison with the geostrophic flow and the Taylor-Proudman balance is restored.
The meridional circulation is dominated by a single, global circulation pattern in each hemisphere.
For positive $b$ (solar-like differential rotation case), the meridional circulation is counter-clockwise and the differential rotation shows an equatorial acceleration as the imposed differential rotation at the boundary.
For negative $b$, i.e. when the equator is slower than the pole, 
we observe the same behavior but the sense of the meridional circulation is clockwise and the field of differential rotation is reversed with a polar acceleration. At the surface, the fluid moves toward the equator which rotates slower than the pole.

With MESA models we explored the influence of the stellar
mass on the foregoing results, but of course remaining in the case
of solar-type stars. In all cases, the differential rotation and
associated meridional circulation turn out to have the same properties as
previously described since the shape of the \BVF profile is the same for
masses in $[0.4-1.4]M_\odot$. The amplitudes of the flows are slightly
different but remain of the same order of magnitude as in the solar case.

\subsection{Core-to-surface rotation ratio}

\begin{figure}
	\begin{center}
\includegraphics[scale=0.5]{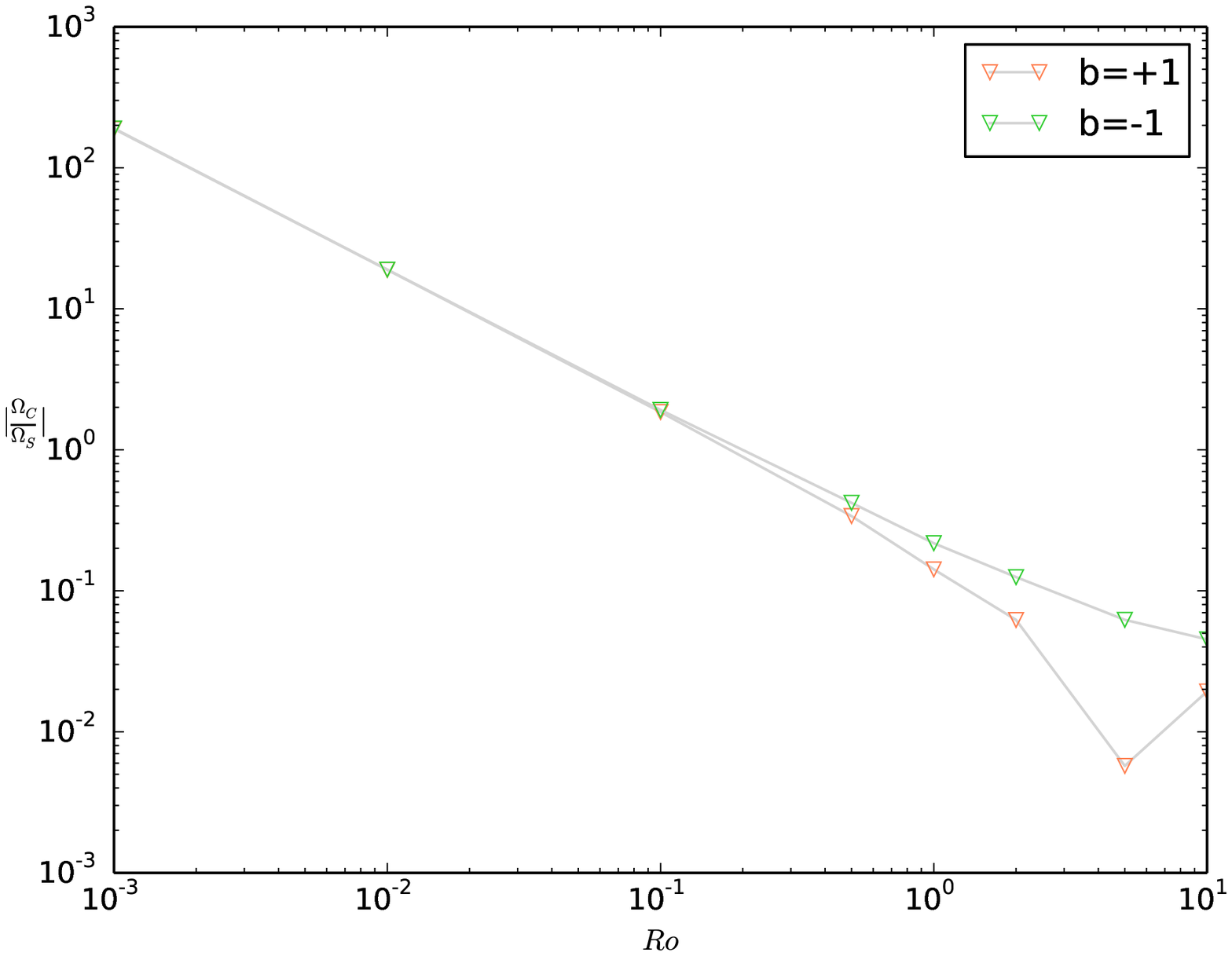}
	\end{center}
\caption{Logarithm of the core-to-surface
rotation ratio versus the logarithm of the Rossby number for $E=2.10^{-6}$.\label{ocs}}
\end{figure}

Asteroseismic analysis provide the core-to-surface rotation ratio of
numerous low-mass stars \citep{benomar15}. Being the only internal
rotation diagnosis as of today because of the low-angular resolution provided by asteroseismology, we therefore compute this ratio as the
latitudinal average of the angular velocity over a radius close to the
center, namely $r_{\rm core} \sim 0.15$, and near the surface, $r_{\rm surf}
\sim 1$

\begin{equation}
\frac{\langle\delta\Omega_{C}\rangle_\theta}{\langle\delta\Omega_{S}\rangle_\theta}=\frac{\int_0^{\pi/2} \delta\Omega(r_{\rm core},\theta)\sin\theta d\theta}{\int_0^{\pi/2} \delta\Omega(r_{\rm surf},\theta)\sin\theta d\theta}\; .
\end{equation}
We do not compute
$\langle\delta\Omega_{C}\rangle_\theta$ at the center $r=0$ since individual
$g$-modes are hardly identifiable in this region \citep{garcia07, appourchaux10, garcia10}. 
When plotting the differential
rotation on Fig. \ref{fig5}, we have subtracted the rotation rate of the
pole at the surface to the rotation rate so that negative value of the
rotation rate means that the examined zone rotates retrogradely in the
reference frame rotating with the pole. The core-to-surface rotation
rate, computed in this framework, is shown in
Fig. \ref{ocs} as a function of the Rossby number
($\Delta\Omega/2\Omega_0$).

The surface shear drives completely the averaged surface rotation rate which has the sign of $b$ and whose amplitude is independent of the Rossby number.
The averaged core rotation rate is strongly impacted by the baroclinic torque amplitude 
which induces a slow shellular-like rotation when compared to the outer latitudinally averaged rotation.
Thus, the averaged core rotation rate is negative in most cases and tends towards zero when increasing $\mathcal{R}o$ (i.e. decreasing the baroclinic torque amplitude).
In conclusion, the core-to-surface rotation ratio (in absolute value) decreases as $\mathcal{R}o$ increases. 
The only departure from this behavior (the small bump for positive $b$ and high $\mathcal{R}o$) is due to a change of sign of the averaged core rotation rate which becomes positive for Rossby number higher than $5$ i.e. when the geostrophic solution completely dominates. In the retrograde case, the baroclinic flow has to have a higher amplitude to overcome the (retrograde) shear-induced flow than in the prograde case. 
The scaling law index is $m=-1$
when writing $|\langle\delta\Omega_{C}\rangle_\theta/\langle\delta\Omega_{S}\rangle_\theta|\propto \mathcal{R}o^m$.
At low Rossby number, when the baroclinic solution is dominant, the core-to-surface rotation ratio is high which means that the differential rotation between the core and the surface is large in absolute value but, when considering the pole rotation, the core is slower than the surface according to our simulation.
Reversely, for large Rossby numbers, the geostrophic solution is dominant and the core-to-surface rotation ratio is small with a rapidly rotating core (respectively slowly) when $b=-1$ (respectively $b=+1$), as illustrated in Fig. \ref{fig5}.

\section{Discussion \& conclusion}
\label{sect3}

\subsection{Rossby number and timescales}

Regarding the 3D numerical simulations results for the differentially rotating convective envelopes of solar-like stars, the relative latitudinal shear has been scaled as follows
\begin{equation}
\frac{\Delta\Omega}{\Omega_0}~\propto~ \Omega_0^m \; ,
\end{equation}
with $m=-0.11$ in the hydrodynamical case or $m=-0.56$ according to MHD numerical simulations \citep{varela16}. This scaling law is closer to the scaling law derived from the observations $\Delta\Omega\propto\Omega^{0.15}$ \citep{barnes05, reinhold13} than the pure hydrodynamical one.
The relative shear, hence the Rossby number in our study, decreases as the stellar rotation rate increases.
For angular velocities between $1$ and $100$ $\Omega_\odot$, which is the range of observed angular velocities in solar-like stars reported in \cite{GB13, GB15}, the corresponding Rossby number is between $7.10^{-2}$ and $5.10^{-3}$ using the MHD scaling law $\Delta\Omega=6.5~10^{-5}\Omega_0^n$ and $n=0.44$ from \citep{varela16}. 
For such values of the Rossby number, our numerical results predict that the dynamics is driven by the baroclinicity with a quasi-shellular differential rotation as illustrated in Fig. \ref{fig5}.
Indeed, we recall that with our setting, the baroclinic torque amplitude increases when the Rossby number decreases.  

But, our study compares the steady solution of the baroclinic and the geostrophic flows.
Meanwhile it is not certain weather these solutions reach their steady states during the main-sequence lifetime (i.e. $10^{10}$ years for a solar-like star).

The characteristic timescale for the settling of the geostrophic solution is the one of a spin-up
\begin{equation}
\tau_{\rm{SU}}=\frac{P}{\sqrt{E}}\; ,
\label{Tau_su}
\end{equation}
where $P$ is the rotation period and $E$ the Ekman number.
In the stellar case, the Ekman number is around $10^{-10}$ leading to $\tau_{\rm{SU}}=10^5P$.
If $P$ is of order tens of days, as is the Sun, the transient solution does not last more than $\sim 10^4$ years, which is short in comparison to the time a star spends on the main-sequence.   

The baroclinic modes damp on the Eddington-Sweet timescale, namely on
\begin{equation}
\tau_{\rm{ES}}=\tau_{\rm{KH}}\frac{N^2}{4\Omega_0^2}\; ,
\end{equation}
where $\tau_{\rm{KH}}=R^2/\kappa\approx 10^8$ years in the solar case, is the thermal Kelvin-Helmholtz timescale.
When the stellar rotation rate increases, the Eddington-Sweet timescale gets shorter and tends to the Kelvin-Helmholtz timescale. This is often the case on the pre-main-sequence.

Hence, in rapid or young rotators the baroclinic steady state is likely to be reached.
For slower (older) rotators like the present Sun, the ratio $\frac{N^2}{4\Omega_0^2}$ is around $10^4$ leading to $\tau_{\rm{ES}}\sim10^{12}$ years. Since it is very large, it means that there are likely residuals of the baroclinic modes (initial conditions).

Therefore, for solar-like stars, the geostrophic solution is most certainly steady while it is not clear for the baroclinic one depending on the stellar rotation rate and thus the age.

We compute the Eddington-Sweet timescale as a function of the angular velocity and find that 
the limit for the steady state to be reached on a solar main-sequence lifetime is $\Omega_0\sim 30 \Omega_\odot$.
Stars rotating faster than $\Omega_0\sim 30 \Omega_\odot$ can thus reach a steady baroclinic state and have a baroclinic dynamics within their radiative zone.
For stars rotating slower than $\Omega_0\sim 30 \Omega_\odot$, the dynamics is not determined
directly because the baroclinic flow is unsteady and depends on initial conditions.

\begin{figure}
	\begin{center}
\includegraphics[scale=0.5]{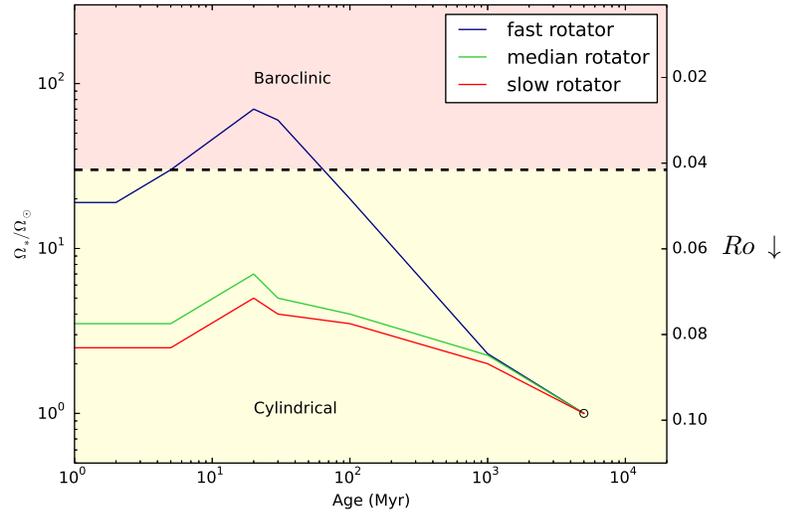}
	\end{center}
\caption{Schematical trends of angular velocity $\Omega_\star/\Omega_\odot$ of the envelope of fast (blue line), median (green line), and slow (red line) solar-type rotators as a function of age from \cite{GB13}. The open circle is the angular velocity of the present Sun. Using the scaling law from \cite{varela16}, the red area delimits where the dynamics should be driven by the baroclinicity leading to a shellular rotation. The yellow area shows where our modelling allows to expect a cylindrical differential rotation in the radiative core of solar-like stars if taking only into account rotation and induced large-scale flows.\label{figurebilanGB}}
\end{figure}
If we consider that the geostrophic flow is steady and that the baroclinic flow amplitude superposes on it according to the following transient time evolution
\begin{equation}
\mathcal{R}o^{-1}\left(\frac{1-e^{-\tau/\tau_{\rm{ES}}}}{1-e^{-1}}\right)\; ,
\end{equation}
its amplitude gets comparable to the one of the geostrophic flow at
\begin{equation}
\tau = -\tau_{\rm{ES}} \ln{[1-\mathcal{R}o(1-e^{-1})]} \; ,
\end{equation}
which is, within the range of $\Omega_0\in[\Omega_\odot,10^2\Omega_\odot]$, always longer or comparable than the main-sequence lifetime because of the amplitude of the Rossby number.
Therefore, the dynamics of the radiative core of solar-type stars rotating slower than $\Omega_0\sim 30 \Omega_\odot$ may be dominated by the shear with a cylindrical differential rotation imposed by the geostrophic balance according to the initial conditions.
These would be the differential rotation profile at the end of the pre-main-sequence when the gravitational contraction ends. \cite{HR14} have shown that it would be also cylindrical. 

We summarize these results in Fig. \ref{figurebilanGB}, where we use the results from \cite{GB13}.
We plot schematical rotational evolution of the envelope of solar-like stars they obtained in a rapid, median and slow rotating case. We delimit the region where we can expect the steady baroclinic solution to drive a shellular differential rotation in red and the region where the differential rotation may be cylindrical due to the imposed shear (yellow region).
The steady state analysis is clearly limited for this study even though we make such an approximation in order to keep the equations linear since the non-linear term is also proportional to $\mathcal{R}o$.
A time evolution analysis of the baroclinic torque over initial conditions of the geostrophic flow would insure this prediction.

If we use the scaling law derived from the observations by \cite{barnes05} and \cite{reinhold13}, we obtain high Rossby numbers 
in such a way that, comparing only the steady state amplitudes, the baroclinic torque would be very small and the cylindrical rotation profile induced by the shear would be dominant for all rotation rates.

Also, on the main-sequence, solar-type stars age is correlated to the rotation rate through a gyrochronological law \citep{skumanich72, kawaler88, revilleX, vansaders16} since stars undergo a spin-down through stellar wind.
Indeed, the wind is responsible for a mass and angular momentum loss expected to generate a spin-down geostrophic solution in the bulk of the star \citep{RB14} on a timescale similar to the spin-up case given by Eq. (\ref{Tau_su}), i.e. a short timescale regarding the duration of the main-sequence. 
Therefore, a complete study would require to also consider the spin-down flow from \cite{RB14} in competition with the baroclinic flow and the flow driven by the imposed convective differential rotation. 

Our simple model predicts that the Sun radiative core should have a cylindrical differential
rotation with an angular velocity gradient that is rather mild and similar to the one deduced from
asteroseismic observations of solar-like ($M\in[1,1.6]M_\odot$) main-sequence stars
\citep{benomar15}. However, the cylindrical differential rotation is in contradiction with the
observed solid-body rotation of the present solar radiative zone (at least for
$r\geq0.2R_\odot$. Other processes therefore come into play like, possibly, internal gravity
waves \citep{zahn97}, hydrodynamic shear-induced anisotropic turbulence \citep{zahn92} or
magnetic fields \citep{mestelweiss87, GM98}.

\subsection{Conclusion}

In order to study the dynamics induced by the shear that a surface convective
envelope imposes on an internal radiative core in main-sequence low-mass
stars, we build a simplified model of a central radiative zone on the top
of which we impose a latitudinal shear boundary condition.  We consider
a simple latitudinal shear at the surface of the model inspired by
heliosismology inversion profiles and global numerical simulations
of the convective envelope of differentially rotating low-mass stars.
This allows us to quantify the impact of the shear between the pole and
the equator of the convective/radiative interface on the steady flow
within the radiative core.

Analytically, we find that the imposed shear drives an $\mathcal{O}(1)$
geostrophic solution in addition to the thermal wind rising from the
stable stratification. 
The baroclinic flow is characterized by a quasi shellular differential rotation with multiple cells of meridional circulation while the geostrophic flow tends to sustain a cylindrical
differential rotation.  
When the geostrophic solution induced by
the shear is dominant (high Rossby number regime $\mathcal{R}o=\Delta\Omega/2\Omega_0>1$), there is a unique cell in each hemisphere which is
quite similar to the previous results obtained by \cite{friedlander76}
and \cite{garaud02} in the solar case. The high Rossby number case must be
interpreted very carefully since it is formally out of the linear regime. Indeed,
for $\mathcal{R}o>1$, the non-linear advection term may be important and we
did not include it in this work.

Since the baroclinic torque amplitude is proportional to the inverse of the Rossby number, 
we evaluate its value according to the scaling laws found in 3D numerical simulations by \cite{varela16} to determine if the spin-up flow from the shear overtakes the baroclinic flow, .
The Rossby number is found to decrease as the global stellar rotation rate increases.
For such values of Rossby numbers ($\mathcal{R}o\sim 10^{-2}$), the baroclinic flow is expected to be dominant for stars rotating faster than $30\Omega_\odot$.
For slower rotating stars, the baroclinic flow is probably still transient and less important in amplitude than the geostrophic flow. Therefore, we expect these stars to have a cylindrical differential rotation.
Scaling laws for Rossby numbers directly derived from observations \citep{reinhold13} suggest that the cylindrical differential rotation profile is dominant for all rotators.
Regarding our 2D models with parameters closest to the Sun,
we do not reproduce the flat rotation profile observed both in radius and latitude in the solar
radiative zone. 
As in previous results obtained in 1D, this strengthens the need to take into account an efficient process
responsible for extra transport of angular momentum.  Such a process could
be internal gravity waves \citep{TC05}, magnetic fields
\citep{mathiszahn05, strugarek11, AAGW13} and the anisotropic turbulent
transport (Mathis et al., 2017, submitted). Their effects will be studied
in forthcoming works.

In addition, the difference in the sign of the differential rotation between \cite{MR06}
models and the compressible ESTER main-sequence models \cite[][]{ELR13}
also suggests that the compressibility may play an important role and that
the Boussinesq approximation  is just a first step before using a
more detailed modelling like one using the anelastic approximation.

The general method presented here will be also applied to other types of stars
such as intermediate-mass and massive stars with a differentially rotating convective core and
$F$-type stars with both a differentially rotating convective core and envelope.

\begin{acknowledgements} 
The authors thank the referee for their comments that allow to improve the article.
D. Hypolite and S. Mathis acknowledge funding by the European Research Council
through ERC SPIRE grant 647383 and the CNES PLATO grant at CEA Saclay.
The authors also wish to thank the MESA website.
M. Rieutord acknowledges funding by the the French Agence Nationale de
la Recherche (ANR), under grant ESRR (ANR-16-CE31-0007-01).  This work
was also supported by the Centre National de la Recherche Scientifique
(CNRS, UMR 5277), through the Programme National de Physique Stellaire
(PNPS), and by the CNES PLATO grant at IRAP.
\end{acknowledgements}

\bibliographystyle{aa}
\bibliography{mabib}

\appendix

\section{Spectral expansion of hydrodynamical equations}
\label{MN2}
The radial grid points, corresponding to the expansion onto a Gauss Lobatto grid derived with Chebyshev polynomials, are defined as

\begin{equation}
\left \{
\begin{array}{l}
r_j=\frac{1}{2}\left(1-\cos(\displaystyle{\frac{j\pi}{N_r-1}})\right)\; ,\\
0 \leq j \leq N_r-1\; ,\\
r_j \in[0;1]\; .
\end{array}
\right .
\end{equation}
The colatitude dependence of the solution is described with the vectorial
spherical harmonics $(\vec{R}_l^m,\vec{S}_l^m,\vec{T}_l^m)$
\cite[e.g.][]{R87}.  We represent the velocity field as follows

\begin{multline}
\vec{u}=\sum \limits_{l=0}^{+\infty} \sum \limits_{m=-l}^{+l} \{u_m^l(r) \vec{R}_l^m(\theta,\varphi)\\
+ v_m^l(r) \vec{S}_l^m(\theta,\varphi) + w_m^l(r) \vec{T}_l^m(\theta,\varphi)\}\; ,
\end{multline}
where
\begin{equation}
\vec{R}_l^m=Y_l^m(\theta,\varphi) \vec{e_r},\qquad \vec{S}_l^m=\vec{\nabla}_HY_l^m, \qquad \vec{T}_l^m=\vec{\nabla}_H\wedge \vec{R}_l^m\; .
\end{equation}
$Y_l^m$ are the normalized spherical harmonics \cite[e.g.][]{cohentannoudji}, $\vec{e_r}$ is the
unit radial vector and the horizontal gradient operator $\vec{\nabla}_H = \partial_{\theta} \vec{e}_\theta + \frac{1}{\sin\theta}\partial_{\varphi}\vec{e}_\varphi$ is defined on the
unit sphere.
The axisymmetry of the solutions imposes $m=0$ ($\partial_\varphi=0$).  For this
reason we will not write the index $m$ in the following.

\subsection{Hydrodynamical equations}

The continuity equation (\ref{eqcontpoura}) reads on this expansion
\begin{equation}
v^l= \frac{1}{l(l+1)} \frac{1}{r}\frac{\partial}{\partial r}(r^2 u^l)\; .
\label{eqcontHS}
\end{equation}
The vorticity equation (\ref{rot_eqd}) projected onto the $\vec{R}_l$ function is written
\begin{multline}
A_{l-1}^l r^{l-1}\frac{\partial}{\partial r} \left ( \frac{u^{l-1}}{r^{l-2}} \right ) +A_{l+1}^l r^{-l-2}\frac{\partial}{\partial r}\left ( r^{l+3}u^{l+1} \right)\\
+ E \Delta_{l} w^{l} = 0\; ,
\end{multline}
and onto the $\vec{T}_l$ direction
\begin{multline}
B_{l-1}^l r^{l-1}\frac{\partial}{\partial r} \left ( \frac{w^{l}}{r^{l-1}} \right ) +B_{l+1}^l r^{-l-2}\frac{\partial}{\partial r} \left ( r^{l+2}w^{l+1} \right )
\\- E \Delta_{l} \Delta_{l}(ru^{l})=\sqrt{\frac{16\pi}{5}}\varepsilon n^{2}(r)\delta_{l2}\; ,
\end{multline}
where $\delta_{ij}$ is the Kronecker symbol and $\Delta_{l}=\frac{1}{r}\frac{\partial^2}{\partial r^2}r -\frac{l(l+1)}{r^2}$.
The projection onto the $\vec{S}_l$ function is redundant with the first one and the component of the velocity field $v^l$ is computed with the continuity equation projection Eq. (\ref{eqcontHS}).
The coupling coefficients read
\begin{equation}
\left \{
\begin{array}{ccl}
A^{l}_{l+1}=\displaystyle{\frac{1}{(l+1)}\frac{1}{\sqrt{(2l+1)(2l+3)}}} \;  ,\\
A^{l}_{l-1}=\displaystyle{\frac{1}{l}\frac{1}{\sqrt{(2l-1)(2l+1)}}} \;  ,\\
B^{l}_{l+1}=\displaystyle{\frac{l(l+1)(l+2)}{\sqrt{(2l+1)(2l+3)}}} \;  ,\\
B^{l}_{l-1}=\displaystyle{\frac{l(l^2-1)}{\sqrt{(2l-1)(2l+1)}}}\; .
\end{array}
\right.
\end{equation}

\subsection{Boundary conditions at the upper boundary}
\label{bc}

The azimuthal velocity written on the $(\vec{R}_l,\vec{S}_l,\vec{T}_l)$ basis reads
\begin{equation}
u_\varphi(r,\theta)=- \sum \limits_{l=1}^{+\infty} w^l(r) \partial_\theta Y_l(\theta)\; .
\end{equation}
Because the model is symmetric with respect to the equator, the azimuthal velocity has the property to be fully described by odd $l$.
Using only the two first harmonics $l=1$ and $l=3$, 
the azimuthal velocity reads
\begin{equation}
\begin{array}{ccl}
u_\varphi(r,\theta)&=&w^{l=1}(r) \sin\theta \sqrt{\frac{3}{4\pi}} \\
 &-& w^{l=3}(r) \left( 4\sin\theta + 5\sin^3\theta\right) \sqrt{\frac{7}{4\pi}} \frac{3}{2}\; .\\
\end{array}
\end{equation}
In order to set the shear described by the expression (\ref{scaledbc}), we then set at the surface
\begin{equation}
w^{l=1}(r=1)=-b\frac{4}{5}\sqrt{\frac{4\pi}{3}}\; ,
\end{equation}
and
\begin{equation}
w^{l=3}(r=1)=-\frac{b}{5}\frac{2}{3}\sqrt{\frac{4\pi}{7}}\; .
\end{equation}
All other $l$-components of the azimuthal velocity are zero at the upper boundary. 
Components with even $l$ numbers ($u^l$, $v^l$) are also zero because we set to zero the meridional velocity components at $r=1$.

\end{document}